# Overcritical states of a superconductor strip in a magnetic environment

Yu. A. Genenko
*Technische Universität Darmstadt, FB Materialwissenschaft, Petersenstraße 23, D-64287 Darmstadt, Germany and Zentrum für Funktionswerkstoffe gGmbH, Windausweg 2, D-37073, Göttingen, Germany*

A. Snezhko
*Faculty of Mathematics and Physics, Charles University, V Holesovickach 2, 180 00 Prague 8, Czech Republic*

H. C. Freyhardt
*Institut für Materialphysik, Universität Göttingen, Hospitalstraße 3-7, D-37073, Göttingen, Germany and Zentrum für Funktionswerkstoffe gGmbH, Windausweg 2, D-37073, Göttingen, Germany*

A current-carrying superconducting strip partly penetrated by magnetic flux and surrounded by a bulk magnet of high permeability is considered. Two types of samples are studied: those with critical current controlled by an edge barrier dominating over the pinning, and those with high pinning-mediated critical current masking the edge barrier. It is shown for both cases that the current distribution in a central flux-free part of the strip is strongly affected by the actual shape of the magnetic surroundings. Explicit analytical solutions for the sheet current and self-field distributions are obtained which show that, depending on the geometry, the effect may suppress the total loss-free transport current of the strip or enhance it by orders of magnitude. The effect depends strongly on the shape of the magnet and its distance to the superconductor but only weakly on the magnetic permeability.

## I. INTRODUCTION

The most important property of superconductors essential for large-scale applications is their ability to carry a large transport current without energy dissipation. In a Meissner state, which is realized at magnetic fields (including a self-field of the current) smaller than the lower critical field of the material, $H_{c1}$, there is no magnetic flux inside a superconducting sample and hence no dissipation, although the field $H_{c1}$ is small for practical low-$T_c$ and very small for high-$T_c$ superconductors which easily facilitates to overcome the flux-free state. Besides, in the Meissner state the current flows only in thin surface layer of a bulk superconductor or mostly along the edges of a superconducting ribbon. This restricts an effective usage of the inner part of superconductor and results in a low average value of a total transport current. Therefore it is generally believed that an appropriate candidate for a high-current nondissipative conductor is a hard type-II superconductor filled with a magnetic flux in the so-called critical state[1] where the magnetic field penetrates the whole sample in the form of Abrikosov magnetic vortices but all vortices are pinned by some pinning centers which prevents the vortex motion and dissipation. Thus the main concern of superconductor developers during the last three decades was an improvement of the material quality (grain interconnections, texture, orientation, etc.) and, at the same time, an introduction of suitable effective pinning centers.

In this work we develop another approach to the improvement of current carrying capability of a superconductor strip, based on the protection of the Meissner state by magnetic screening.[2] Two sorts of specimens are considered: those possessing the large edge barrier against magnetic vortex entry which dominates over the pinning, and those possessing strong pinning masking completely the edge barrier.

For the first specimens, the entirely flux-free Meissner state of magnetically shielded superconductor strips is studied. The nature of the edge barrier itself (Bean-Livingston,[3,4] geometrical,[5–7] or enhanced bulk pinning at the edges[8,9]) does not matter. In what follows, we imply a most robust geometrical barrier. For the second sort of samples, we consider by means of critical state model the magnetically shielded strips partly penetrated by magnetic flux.

It is shown that, in both cases, current distributions in flux-free regions of strips are very sensitive to presence of soft magnets of high permeability. Depending on the form of magnetic environment the total loss-free transport current may be strongly suppressed or enhanced.

Let us note here that, on the contrary to a bulk case, an isolated hard superconductor sheet partly filled with a magnetic flux carries an essential part of a current in the vortex-free inner region.[9–13] As was stressed by Zeldov *et al.*,[13] at a total transport current of $0.6I_c$, where $I_c$ is a total critical current of the sheet, 2/3 of the current flows in the vortex-free region. We show in this work that the partly flux-free sheet influenced by controlled external conditions may carry the total current considerably exceeding $I_c$.

In Sec. II, we show that a superconducting sheet surrounded by a soft magnetic material exhibits a very unusual current distribution in the Meissner state which may be controlled by magnet shape and its permeability $\mu$. Current distributions over the sheet located near a bulk flat magnet or inserted inside a cylindrical cavity are calculated analytically for $\mu \gg 1$ and numerically for arbitrary magnetic permeabilities.

The Meissner state of strips in more complicated magnetic surroundings is studied in Sec. III by means of the method of images and of conformal mapping which is valid for the case of $\mu\rightarrow\infty$. It is shown that a sheet current, fixed at the sheet edges by the edge barrier, may be drastically enhanced in the inner part of the sheet by a special choice of the shape of the surrounding magnet.

The field and current distributions in magnetically shielded hard superconductor strips partly filled with magnetic flux in the critical state are studied in Sec. IV for the case of flat bulk magnets. The same problem for a convex magnetic cavity is treated in Sec. V by means of conformal mapping of the above-considered flat surrounding. In particular, it is shown the possibility of stable overcritical sheet current distributions resulting in the large enhancement of the total transport current. The results are summarized and discussed in Sec. VI.

## II. CURRENT REDISTRIBUTION IN A FLUX-FREE STRIP BY A MAGNET OF ARBITRARY MAGNETIC PERMEABILITY

In this section we consider the rearrangement of the Meissner state of a superconductor strip in simple configurations of a magnetic surrounding which may be simply treated by the method of images.[14,15] To remind the reader general results relevant to our problem we consider first the virgin state of an isolated superconductor sheet.[11,12,16,17]

### A. Virgin state of the isolated superconductor strip

Let an infinitely long flat strip occupy the space $|y|\leqslant d/2, |x|\leqslant W/2$ and carry a transport current $I$ in positive $z$ direction. All physical quantities are $z$ independent and $z$ component of the magnetic self-field of the current equals zero. We consider the case of a wide strip of width $W$ implying the $W\gg\delta=\max(\Lambda,d)$ where $\Lambda=\lambda^2/d$ is the transverse penetration depth relevant to very thin films with $d<\lambda$ (Ref. 18) and $\lambda$ is the magnetic-field penetration depth for the bulk material. For the present consideration, it is irrelevant whether the thickness $d$ is larger or smaller than $\lambda$.

The electrodynamics of such a strip may be described to the accuracy of $\delta/W$ by a sheet current[11,12]

$$J(x)=\int_{-d/2}^{d/2}dy\, j(x,y), \tag{1}$$

where $j(x,y)$ is a local current density in the positive $z$ direction. The total value of current is fixed by the condition

$$\frac{W}{2}\int_{-W/2}^{W/2}du\, J(u)=I. \tag{2}$$

Leaving aside the details on the scale of $\delta$, the magnetic-field distribution around the strip in the $(x,y)$ plane is given by a superposition of elementary current contributions $J(u)du$ according to the Ampere law[11,12]

$$\mathbf{H}(x,y)=\frac{1}{2\pi}\int_{-W/2}^{W/2}du\, J(u)\frac{(-y,x-u)}{(x-u)^2+y^2}. \tag{3}$$

Using the formula $y/[y^2+(x-u)^2]\rightarrow\pi\delta(x-u)$ for $y\rightarrow+0$ one easily relates a sheet current to a jump of the parallel field component at the sheet plane

$$J(x)=H_x(x,y=-0)-H_x(x,y=+0). \tag{4}$$

Contrary to the parallel component, the perpendicular component $H_y$ does not change essentially within the strip along the $y$ axis and thus $H_y(x,\pm d/2)\simeq H_y(x,0)=H(x)$. To the same accuracy, one finds

$$H(x)=\frac{1}{2\pi}\int_{-W/2}^{W/2}\frac{du\, J(u)}{x-u}. \tag{5}$$

The Meissner flux-free state of the sheet means the absence of magnetic flux inside it and hence the vanishing of the $H_y$ component of the field at the sheet surface. This gives the condition of the virgin state

$$\int_{-W/2}^{W/2}\frac{du\, J(u)}{x-u}=0, \quad |x|<W/2. \tag{6}$$

In the above formulas (4)–(6), the sheet is treated as one of zero thickness[18] with the same accuracy as Eq. (3).

The solution to Eq. (6) corresponding to the total current $I$ given in[16,17]

$$J(x)=\frac{I}{\pi\sqrt{(W/2)^2-x^2}} \tag{7}$$

diverges formally at the sheet edges where it, in fact, saturates on the scale of $\delta$ and achieves a maximum value of $J_\Gamma=I/\pi\sqrt{W\delta}$. If $J_\Gamma$ is smaller than a critical value $J_b$ determined by some edge barrier mechanism[4–9] or critical current $J_c$ determined by pinning strength[1] the flux-free state may exist. If $J_\Gamma$ exceeds $J_c$ and $J_b$ the flux enters the sample and form a profile determined by the pinning strength inside.[1]

We consider in this section the samples with barrier-controlled critical currents which means $J_b>J_c$. For example, the geometrical barrier mechanism gives a thickness independent $J_b=2\varepsilon_0/\Phi_0$,[5,6] where $\varepsilon_0=\Phi_0 H_{c1}$ is the material dependent energy per unit length of a magnetic vortex and $\Phi_0$ is the unit flux quantum. For a wide thin YBCO film of thickness $d=50$ nm and width $W=5$ mm at 77 K one may expect $\lambda=0.5$ $\mu$m,[19] $\Lambda=5$ $\mu$m, $\delta=2\times10^{-3}$, and $J_b=0.52\times10^4$ A/m. Then the maximum average current density still saving the Meissner state $j_M=I/Wd=(J_b/d)\pi\sqrt{\delta/2}$ would amount to $1.02\times10^6$ A/cm$^2$.

This value may easily be masked in epitaxial YBCO films of good quality which possess often higher pinning-mediated critical currents[20] unless special efforts are taken to obtain low-pinning samples and to refine the effect of a possible edge barrier.[21] Better opportunities for the investigation of the effect are provided by BSCCO single crystals which possess initially low pinning and which exhibit barrier-mediated critical currents.[5,6] In this case, a suppression of the current peaks at the strip edges through some redistribution of the current in favor of the inner part of the strip would enable the latter to carry a higher transport current in the Meissner state and to compete with the flux-filled critical state of hard superconductors.

### B. Flat magnetic screen

Let us consider the Meissner state of the same superconducting strip in presence of a bulk magnet. Let the magnet fill the half space $x \geqslant a > W/2$ as is shown in Fig. 1. To find the field and current distribution one can solve the Poisson equation for a vector potential $\mathbf{A}=[0,0,A_i(x,y)]$ in both media,[14,15]

$$\Delta A_i = -\mu_i J(x)\delta(y), \tag{8}$$

where $\mu_i=\mu_1$ and $\mu_i=\mu_2$ are the magnetic permeabilities of the media at $x<a$ and $x \geqslant a$, respectively, and the sheet current $J(x)$ is defined within the interval $(-W/2,W/2)$. The vector potentials $A_i$ in both media and their derivatives normal to the surface obey the boundary conditions at the surface of the magnet,

$$A_1 = A_2 + \text{const}, \quad \frac{1}{\mu_1}\frac{\partial A_1}{\partial n} = \frac{1}{\mu_2}\frac{\partial A_2}{\partial n}, \quad x=a. \tag{9}$$

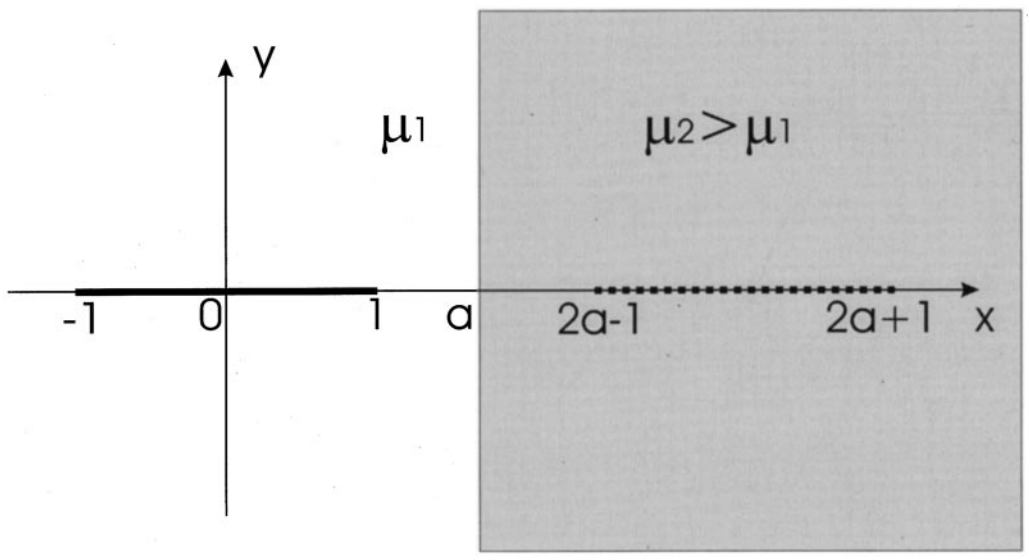

FIG. 1. Cross section of a current-carrying sheet near the boundary of a magnetic half space. The sheet and its image (shown with dotted line) lie in the plane perpendicular to the magnet boundary.

Magnetic inductions in both media are connected to the $\mathbf{A}_i$ by $\mathbf{B}_i = \text{curl}\,\mathbf{A}_i$ and magnetic fields $\mathbf{H}_i = \mu_i^{-1}\mathbf{B}_i$.

The solution to Eq. (8) for an arbitrary sheet current $J(x)$ satisfying the above boundary conditions may be constructed as a superposition of the vector potentials of the original current-carrying strip and its image at the boundary $x=a$[14,15] (Fig. 1). By this procedure, every elementary straight wire located at the line $(u,0)$ and carrying a current $J(u)du$ generates an image located at the line $(2a-u,0)$. Then, using Ampere's law one finds the vector potentials in both media,

$$A_1(x,y) = -\frac{\mu_1}{4\pi}\int_{-W/2}^{W/2} du\, J(u)\{\ln[y^2+(x-u)^2] + q\ln[y^2+(2a-x-u)^2]\}, \quad x \leqslant a,$$

$$A_2(x,y) = -\frac{\mu_2(1-q)}{4\pi}\int_{-W/2}^{W/2} du\, J(u)\ln[y^2+(x-u)^2], \tag{10}$$

where the parameter $q=(\mu_2-\mu_1)/(\mu_2+\mu_1)<1$ is the image current of the unit current.

The magnetic field around the strip in medium 1 is

$$\mathbf{H}(x,y) = \frac{1}{2\pi}\int_{-W/2}^{W/2} du\, J(u)\left[\frac{(-y,x-u)}{(x-u)^2+y^2} + q\frac{(-y,x+u-2a)}{(x+u-2a)^2+y^2}\right] \tag{11}$$

and in medium 2

$$\mathbf{H}(x,y) = \frac{1-q}{2\pi}\int_{-W/2}^{W/2} du\, J(u)\frac{(-y,x-u)}{(x-u)^2+y^2}. \tag{12}$$

As is seen from the latter equation, the magnetic field dies off slowly inside a magnetic medium as $H_y \sim \mu_1 I/\pi(\mu_1+\mu_2)x$ at $x \gg 1$, which should be expected from Ampere's law. This suggests that, under experimental conditions, the effect of current and field redistribution may be noticeable only if the thickness of the magnetic wall is several times larger than the sheet width $W$.

Equating the $H_y$ component of the field at the sheet to zero one finds a condition for the Meissner state

$$\int_{-W/2}^{W/2} du\, J(u)\left(\frac{1}{x-u} + q\frac{1}{x+u-2a}\right) = 0, \tag{13}$$

for all $x \in (-W/2,W/2)$. Generally, this singular integral equation needs a numerical treatment but for the particular case of very high permeability $\mu_2 \gg \mu_1 (q \to 1)$ the analytical solution can easily be found. Upon the substitution of a new variable $v=(u-a)^2$ and of an unknown function $\phi(v) = J(u)/(u-a)$ into Eq. (13), it is reduced to an equation similar to Eq. (6) which leads to a solution

$$J(x) = \frac{2I(a-x)/\pi}{\sqrt{(W^2/4-x^2)(2a+W/2-x)(2a-W/2-x)}}. \tag{14}$$

The above solution may be tested in the physically clear situation of the direct contact of the sheet with the boundary of the magnet $(a \to W/2)$. In this case, the current sheet and its image form together a continuous current sheet lying at $-1 \leqslant x \leqslant 3$. The Meissner-state condition $H_y=0$ at the left physical half of this extended sheet is provided by a properly scaled Eq. (7) and reads

$$J(x) = \frac{2I}{\pi\sqrt{(W/2+x)(3W/2-x)}}, \quad -W/2 \leqslant x \leqslant W/2. \tag{15}$$

One can easily check that Eq. (14) coincides with Eq. (15) for $a=W/2$. Another obvious solution to Eq. (13) is given by Eq. (7) for the case of uniform medium $(\mu_2=\mu_1)$ for which $q=0$.

As may be seen from Eq. (14), the closer the current sheet is to the magnet surface the more suppressed is the current peak at the corresponding side of the sheet. The direct contact to the bulk magnet $(a=W/2)$ of infinite permeability $(q=1)$ suppresses the current peak at the contact edge completely [see Eq. (15)]. Allowing a finite permeability $\mu<\infty$ or finite distance to the sheet $a-1>0$ should restore the peak at least partly.

Equation (13) is not convenient for the numerical study for arbitrary values of parameters $q$ and $a$ because of the singularity it contains. To get rid of this, one can transform Eq. (13) in some Fredholm integral equation with a weak singularity.[22,23] To perform this we use here the invertion procedure suggested by Brandt in Refs. 11 and 12. According to this, the sheet current $J(x)$ may be expressed through the magnetic field $H(x)$, it generates on the sheet, as

$$J(x)=\frac{2I/\pi}{\sqrt{W^2-4x^2}}-\frac{2}{\pi}\int_{-W/2}^{W/2}du\sqrt{\frac{W^2-4u^2}{W^2-4x^2}}\frac{H(u)}{x-u}. \tag{16}$$

Substituting the second term of Eq. (13) for the $H(u)$ in Eq. (16) and performing the integration over $u$ one finds for $J(x)$ the following Fredholm equation with a nonsingular kernel:

$$J(x)=\frac{2I(1+q)}{\pi\sqrt{W^2-4x^2}}-\frac{2q/\pi}{\sqrt{W^2-4x^2}}\int_{-W/2}^{W/2}du\frac{J(u)}{2a-u-x}\sqrt{(2a-x)^2-\frac{W^2}{4}}. \tag{17}$$

One can check that Eq. (17) is satisfied by the solution (15) at $a=W/2, q=1$ and by the solution (7) at $q=0(\mu_1=\mu_2)$ or at $a\to\infty$.

At $a>W/2$, Eq. (17) presents no difficulties for a numerical study. The set of the solutions to Eq. (17) at $2a/W=1$, 1.01, 1.1, 2, and $\mu_1=1$, $\mu_2=\mu=3$, 9, 200, and $\infty$ corresponding to the $q=0.5$, 0.8, 0.99, and 1, respectively, is presented in Fig. 2. It clearly exhibits the suppression of the current peak at the strip edge when it approaches the magnet boundary at any fixed $\mu$ or when $\mu$ grows at any fixed distance $a$.

To compare the effectiveness of the above two measures for the reduction of the sheet current at the edge we plot below the dependence of the current peak value versus $q$ at various fixed $a$, and also the peak value versus $a$ at various fixed $q$ (Fig. 3). One observes a rather pronounced dependence of the edge current on the distance between the sheet edge and the magnet at any fixed $\mu$ [Fig. 3(b)]. On the contrary, the dependence on $q$ is practically linear within the interval $0.5<q<1(3<\mu<\infty)$ at any fixed distance $a$ [Fig. 3(a)].

It is remarkable that the suppression effect has no peculiarities at $q\to 1$. A relative suppression coefficient

$$\nu=\frac{J_{edge}(\mu=1)-J_{edge}(\mu>1)}{J_{edge}(\mu=1)-J_{edge}(\mu\to\infty)} \tag{18}$$

of 97% is achieved already at $\mu=40$ and $\nu=99\%$ at $\mu=200$ [the estimations were made using the sheet current values at the point $2x/W=0.9996$ plotted for $2a/W=1.01$ in Fig. 3(a)]. Thus the case $\mu\to\infty$ studied in Ref. 2 is representative for the current redistribution effect at moderately high $\mu$'s reasonable for many soft ferromagnets at low temperatures.

On the other hand, the distance to magnetic media proves to be crucial for the current distribution over the sheet. It is close to that of the isolated sheet [Eq. (7)] at $2a/W=2$ but changes strongly at $2a/W=1.01$.

All of this allows one to study the influence of the shape of a bulk magnet and its distance to the sheet under the simplifying assumption $q\to 1(\mu\to\infty)$ keeping the main features of the current redistribution. This will be performed below in Sec. III.

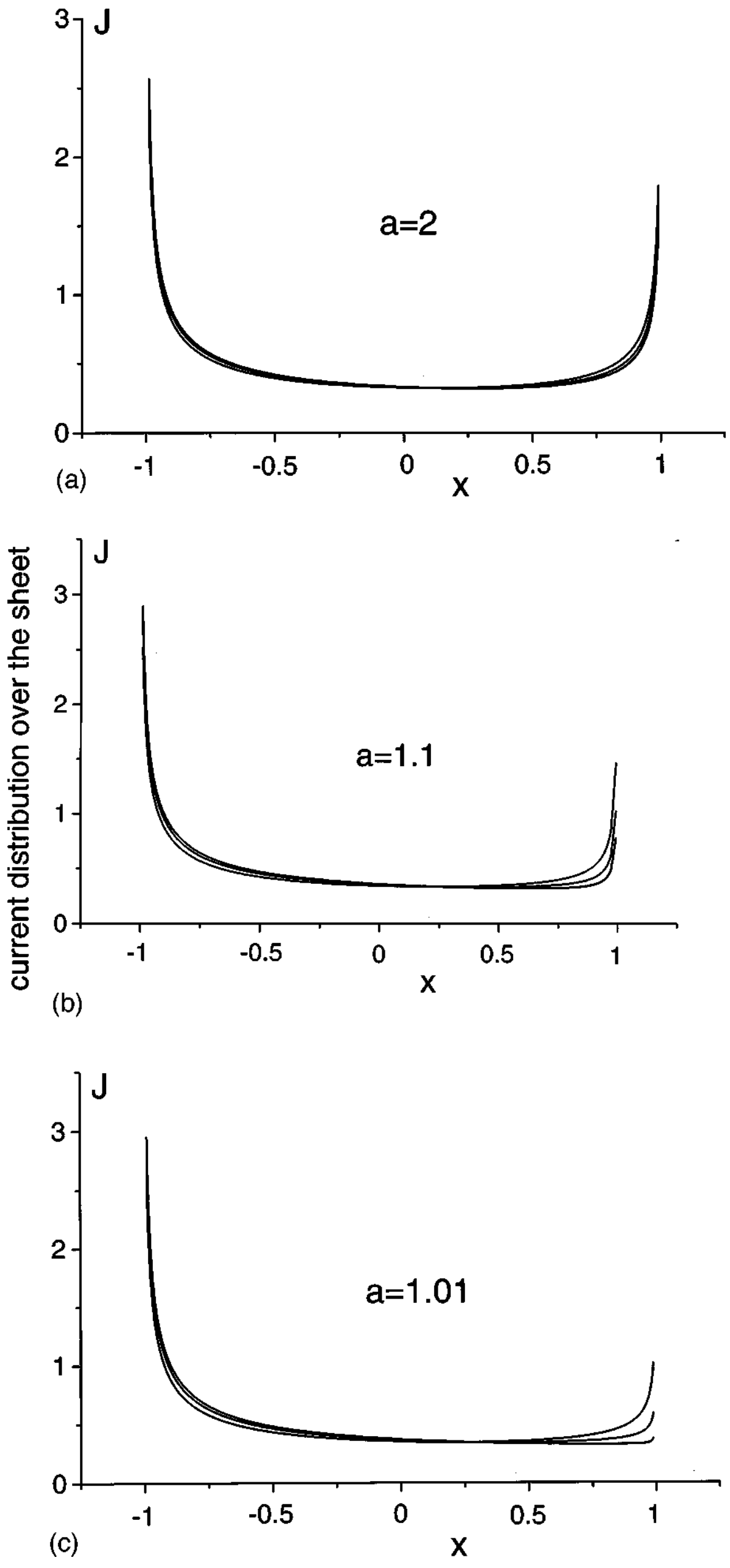

FIG. 2. Distributions of the total transport current $I=1$ over the sheet of width $W=2$ in the case of one-sided screening for various values of distance from the sheet center to the magnet surface (in units of the sheet half-width) $a=2$ (a), 1.1 (b), and 1.01 (c). Different curves in every plot correspond to the magnetic permeabilities $\mu=3$, 9, 200, and $\infty$ ($q=0.5$, 0.8, 0.99, and 1, respectively) and are shown for the in-plane coordinate range $-0.99\leqslant x\leqslant 0.99$. The larger is $\mu$ the more effective is peak suppression at $x=0.99$. The curve for $q=1$ corresponding to the analytical solution (14) cannot be distinguished from that for $q=0.99$.

Of course, the Meissner flux-free state may be saved at high transport currents only if the current peaks are suppressed below the barrier-determined value $J_b$ at both edges of the strip. For this purpose, the current distribution over the sheet in presence of two symmetrical bulk magnets located at $x\leqslant -a$ and $x\geqslant a$, respectively, was considered in Ref. 2. In this case, the original current sheet generates an infinite succession of the images centered at the points $x_n=2an$ of the $x$ axis. Therefore the Meissner state equation $H_y(x,0)=0$ is described in this case by the homogeneous integral equation with a kernel presented by a series in $q^{|n|}$.[2]

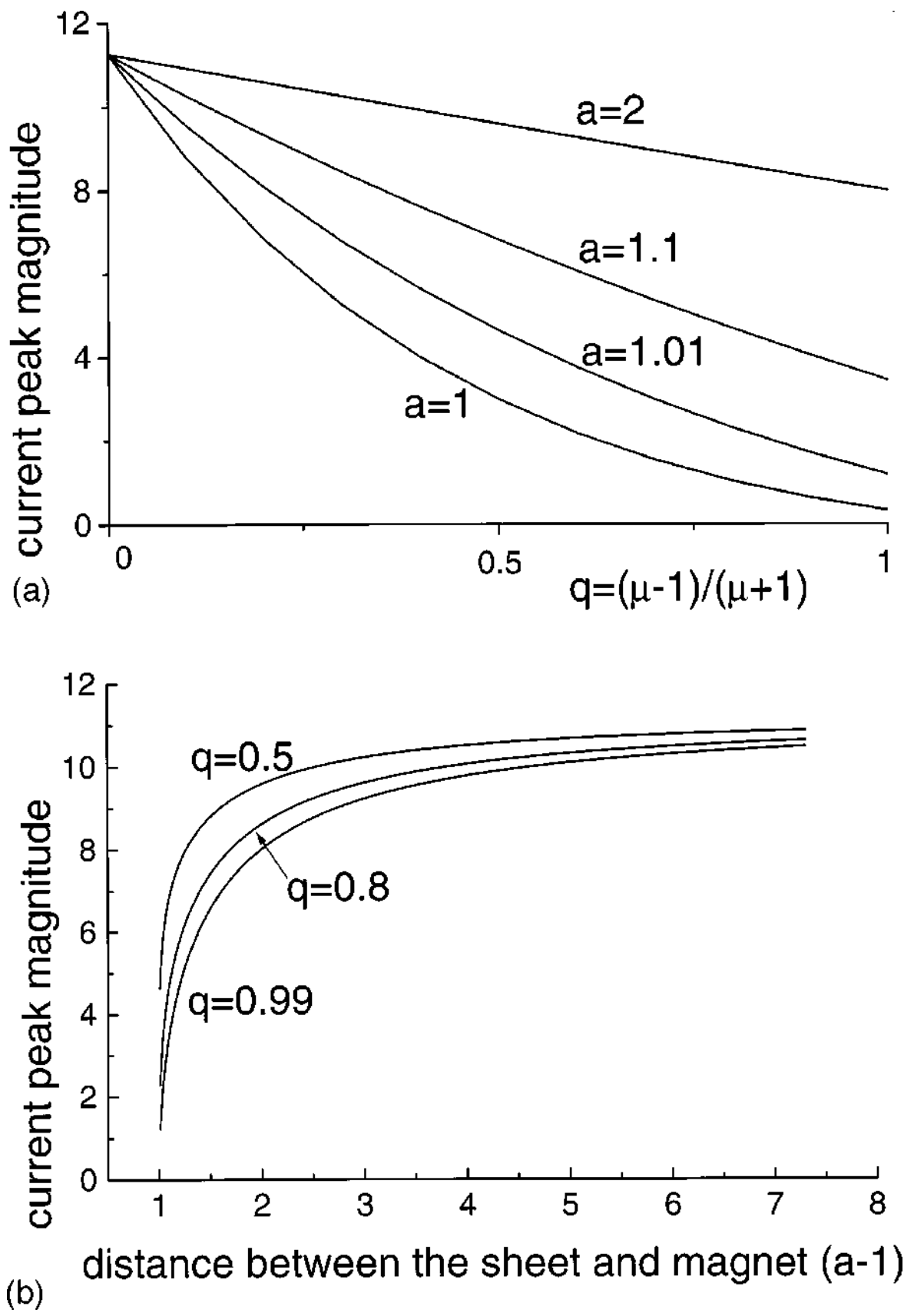

FIG. 3. (a) Dependence of the current peak value taken at the right-hand side of the sheet at $x=0.9996$ in case of one-sided screening versus the image current strength $q$. Different curves correspond to sheet distances to the magnet $a=2$, 1.1, 1.01, and 1 (direct contact). The total current $I=1$. (b) The same current peak versus distance $a$ for different $q=0.5$, 0.8, and 0.99 ($\mu=3$, 9, and 200, respectively).

The solution to this problem was found in Ref. 2 analytically for the case $q=1$ only, when the integral kernel reduced to a circular function. At $0<q<1$, even the numerical study is complicated by the kernel expressed in extended hypergeometric functions. This study is still in progress. Nevertheless, based on the above-performed study of the one-side screening one can expect that the redistribution of the current by the two-sided screening at finite $\mu$ follows the general pattern of the current peak suppression at the strip edges and that the difference between the cases $\mu\to\infty$ and $\mu\simeq 100$ is negligible as well. This understanding is confirmed by considering the case of a symmetrical screening in the next section.

### C. Screening by a thick cylindrical magnetic shell

Consider another case which can be treated in terms of images: A current-carrying sheet inside a cylindrical cavity in a soft magnetic medium (Fig. 4). This configuration might be of practical interest for the usage of a magnetic screening but needs a careful study because of its controversial features. Indeed, it is not clear at a first sight which effect would produce such a surrounding on the current distribution. As was established in Ref. 2, bulk magnets parallel to the strip push the current to the edges of the strip, while bulk magnets perpendicular to the strip result in the flattening of the current distribution. In the case of closed symmetrical surrounding shown in Fig. 4 the above two effects oppose each other.

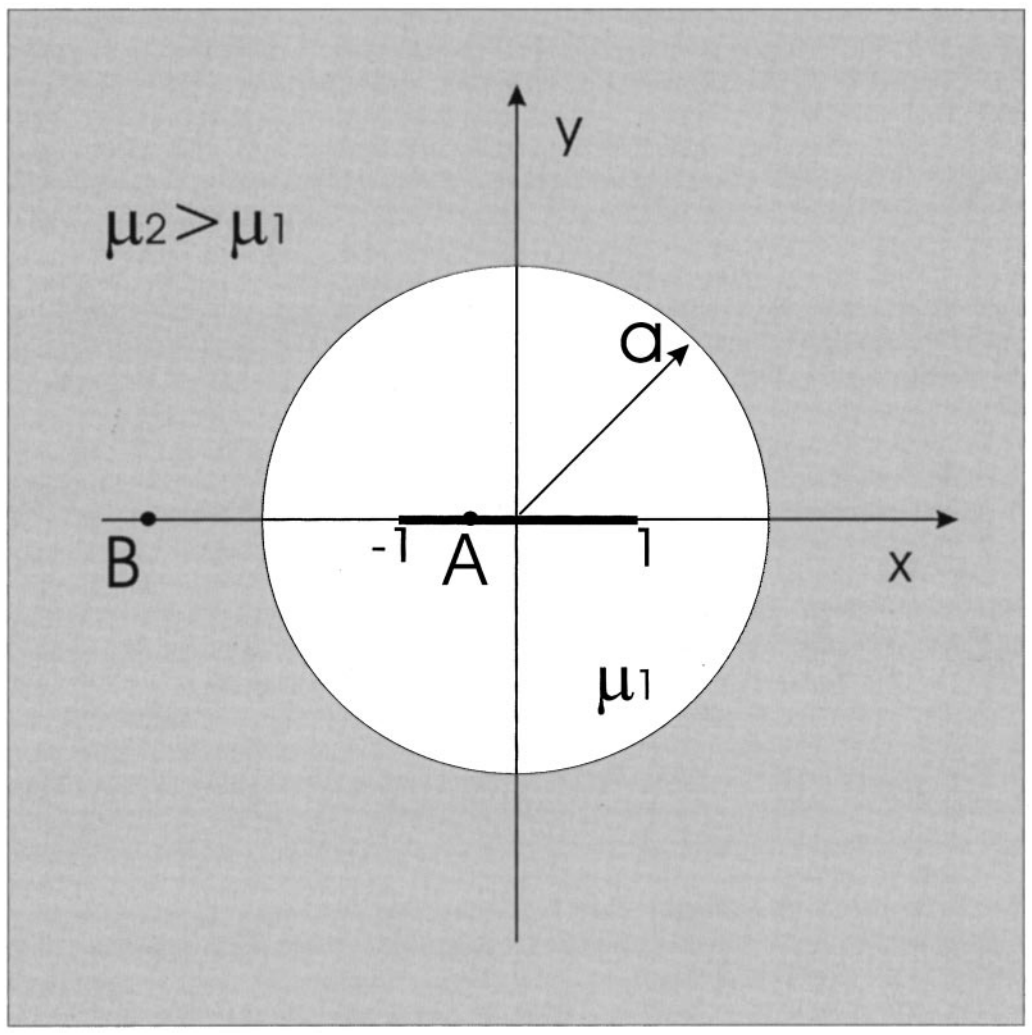

FIG. 4. The solution of the magnetostatic problem of a current sheet within a cylindrical magnetic cavity may be found by the method of images for arbitrary $\mu$'s. Every elementary current at a position $\mathbf{A}(u,0)$ produces an image at a position $\mathbf{B}(a^2/u,0)$.

Let the superconductor strip occupy the same position as in previous sections and be surrounded by a magnet of permeability $\mu_2$ filling the whole space at $x^2+y^2\geqslant a^2$. The magnetic permeability of the medium inside the cavity is $\mu_1$. This problem is also a two-dimensional one and may be treated in terms of a vector potential $\mathbf{A}=[0,0,A(x,y)]$.

Using an analogy to electrostatics[14,15] one can find for every straight current $\tau$ flowing parallel to the $z$ axis along the line $(x=u,y=0)$ (point $A$ in Fig. 4) an image current $q\tau$ flowing the same direction and located at the line $(a^2/u,0)$ (point $B$ in Fig. 4). The vector potentials $A_1$ and $A_2$ in the corresponding media which match at the cavity surface of radius $a$ and which satisfy the condition (9) for their normal to the surface derivatives may be constructed as follows. The vector potential $A_1$ is chosen as superposition of contributions from the current $\tau$ at point $A$ and current $q\tau$ at point $B$. The vector potential in the medium 2 is chosen as superposition of currents $(1-q)\tau$ at a point $A$ and $q\tau$ at the center of the cavity. Then, the total vector potential produced by the current sheet in the magnetic cavity may be found by taking $\tau=J(u)du$ for the elementary current and integration over the sheet width:

$$A_1(x,y)=-\frac{\mu_1}{4\pi}\int_{-W/2}^{W/2}du\,J(u)\{\ln[y^2+(x-u)^2]+q\ln[y^2+(x-a^2/u)^2]\},$$

$$x^2+y^2\leqslant a^2. \tag{19}$$

Inside the magnetic medium

$$A_2(x,y)=-\frac{\mu_2}{4\pi}\int_{-W/2}^{W/2}du\,J(u)\{q\ln(y^2+x^2)$$
$$+(1-q)\ln[y^2+(x-u)^2]\},$$
$$x^2+y^2\geqslant a^2. \tag{20}$$

Then, one can find the magnetic field in the cavity

$$\mathbf{H}(x,y)=\frac{1}{2\pi}\int_{-W/2}^{W/2}du\,J(u)\left[\frac{(-y,x-u)}{(x-u)^2+y^2}\right.$$
$$\left.+q\frac{(-y,x-a^2/u)}{(x-a^2/u)^2+y^2}\right] \tag{21}$$

and in the magnetic medium 2

$$\mathbf{H}(x,y)=\frac{1}{2\pi}\int_{-W/2}^{W/2}du\,J(u)\left[q\frac{(-y,x)}{x^2+y^2}\right.$$
$$\left.+(1-q)\frac{(-y,x-u)}{(x-u)^2+y^2}\right]. \tag{22}$$

The equation of the Meissner state $H_y(x,0)=0$ now reads

$$\int_{-W/2}^{W/2}du\,J(u)\left(\frac{1}{x-u}+q\frac{1}{x-a^2/u}\right)=0 \tag{23}$$

for all $x\in(-W/2,W/2)$. As well as Eq. (13), this equation is singular in the integration domain and may be transformed to the nonsingular Fredholm equation by the invertion procedure. Using as above Eq. (16) and substituting the second term of Eq. (23) for the $H(u)$ one obtains

$$J(x)=\frac{I(1+q)/\pi}{\sqrt{\frac{W^2}{4}-x^2}}-\frac{q/\pi}{\sqrt{\frac{W^2}{4}-x^2}}\int_{-W/2}^{W/2}du\,J(u)\frac{\sqrt{a^4-u^2}}{a^2-ux}. \tag{24}$$

Numerical solutions of Eq. (24) are presented in Fig. 5. Each set of solutions corresponds to a fixed radius of the cavity $2a/W=1.01$, 1.1, and 2 while each curve presents a solution for a definite value of parameter $q$ of the set $q=0.5$, 0.8, and 0.99. One can observe a positive effect of the current peaks reduction at the strip edges by a decreasing $a$ or an increasing $q$. At the same time, the effect of the peak suppression is weaker then that for the case of the flat wall studied above in Sec. II B. The general pattern of the peak behavior when the parameters $a$ and $q$ change remains the same as in the case of the flat wall. The current peaks decrease practically linearly with growing $q$ at any fixed radius $a$ [Fig. 6(a)] and are reduced abruptly as the radius of the cavity decreases down to sheet half width, i.e., $a\to W/2$ [Fig. 6(b)]. The amplitude of the peaks at any fixed $a$ and $q$ are larger than the corresponding magnitudes for the flat wall case which makes the cylindrical cavity less favorable configuration for the purpose of current redistribution by magnetic screening.

One additional interesting case of magnetic screening may be studied with the help of Eqs. (23) and (24) by choosing the negative parameter $q\in(-1,0)$. It presents the situation of superconducting sheet immersed in a magnetic cylinder of permeability $\mu_1$ exceeding the magnetic permeability of the environment $\mu_2$. This configuration reminds the case of a current sheet between bulk magnets parallel to its surface calculated in Ref. 2 and is expected to push the current out to the edges of the sheet.

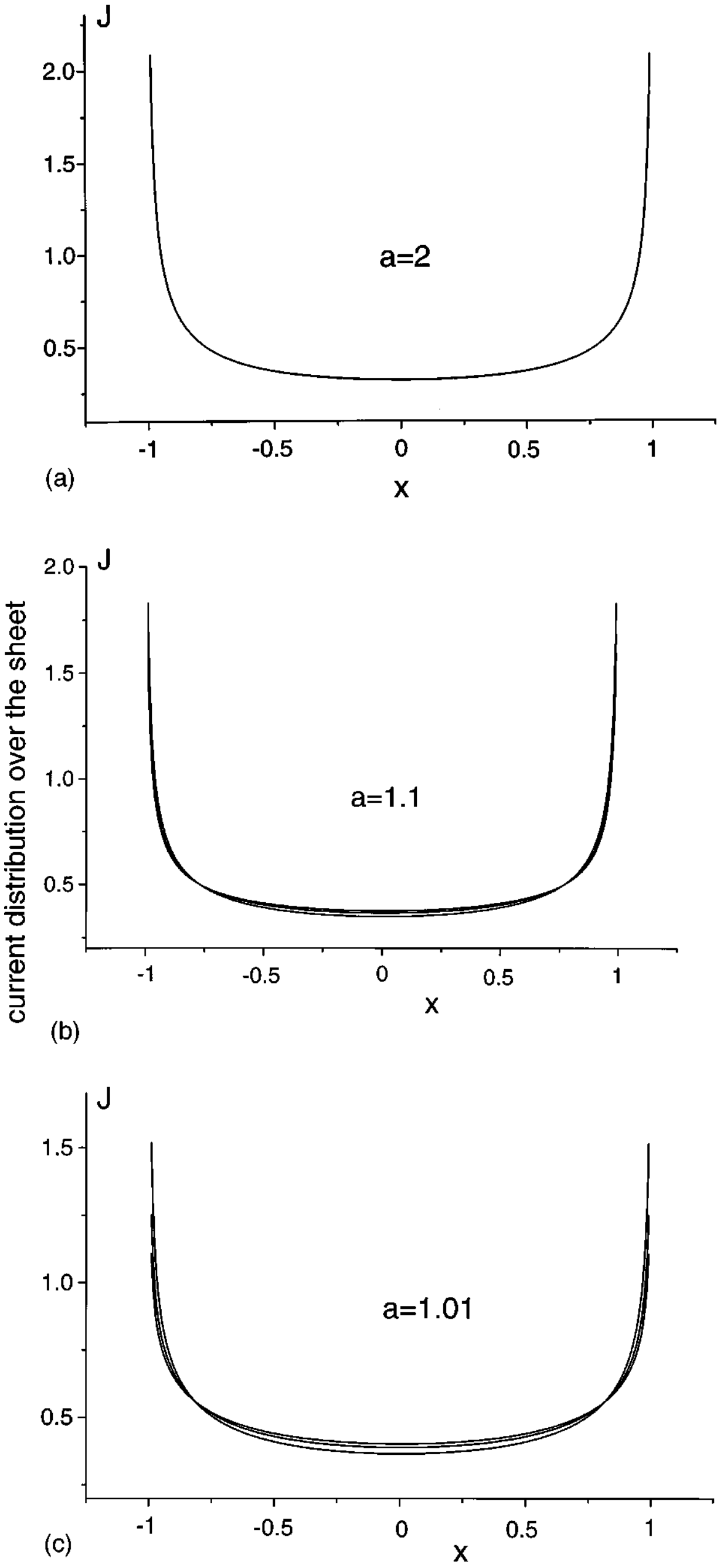

FIG. 5. Transport current distribution over a sheet carrying a total current $I=1$ inside a cylindrical magnetic cavity for various radii of the cavity in units of the sheet half-width $a=2$ (a), 1.1 (b), and 1.01 (c). Different curves in every plot correspond to the magnetic permeability $\mu=3$, 9, and 200 ($q=0.5$, 0.8, and 0.99, respectively) and are shown for the in-plane coordinate range $-0.99\leqslant x\leqslant 0.99$. As well as in Fig. 2 the most effective peak suppression is achieved for the largest $\mu$.

Numerical solution indeed demonstrates such a behavior enhancing the current peaks at the sheet edges in comparison with an isolated sheet. Thus this configuration may not be used for the purpose of improving the current-carrying capability of strips. Nevertheless, it exhibits a nontrivial behavior shown in Fig. 7. The magnitude of the current peak (at the point $2x/W=0.9996$) plotted against a parameter $q$ turns out to have a well defined maximum for all values of the cylinder radius $a$. The maximum remains (though shifted) also at any other choice of the edge point chosen to cut the divergence of the sheet current. This means that the current distribution over the sheet covered with the thick magnetic shell behaves nonmonotonously as a function of the permeability $\mu$ in contrast to the current peak dependence on $\mu$ for sheets in magnetic cavities [Figs. 3(a) and 6(a)].

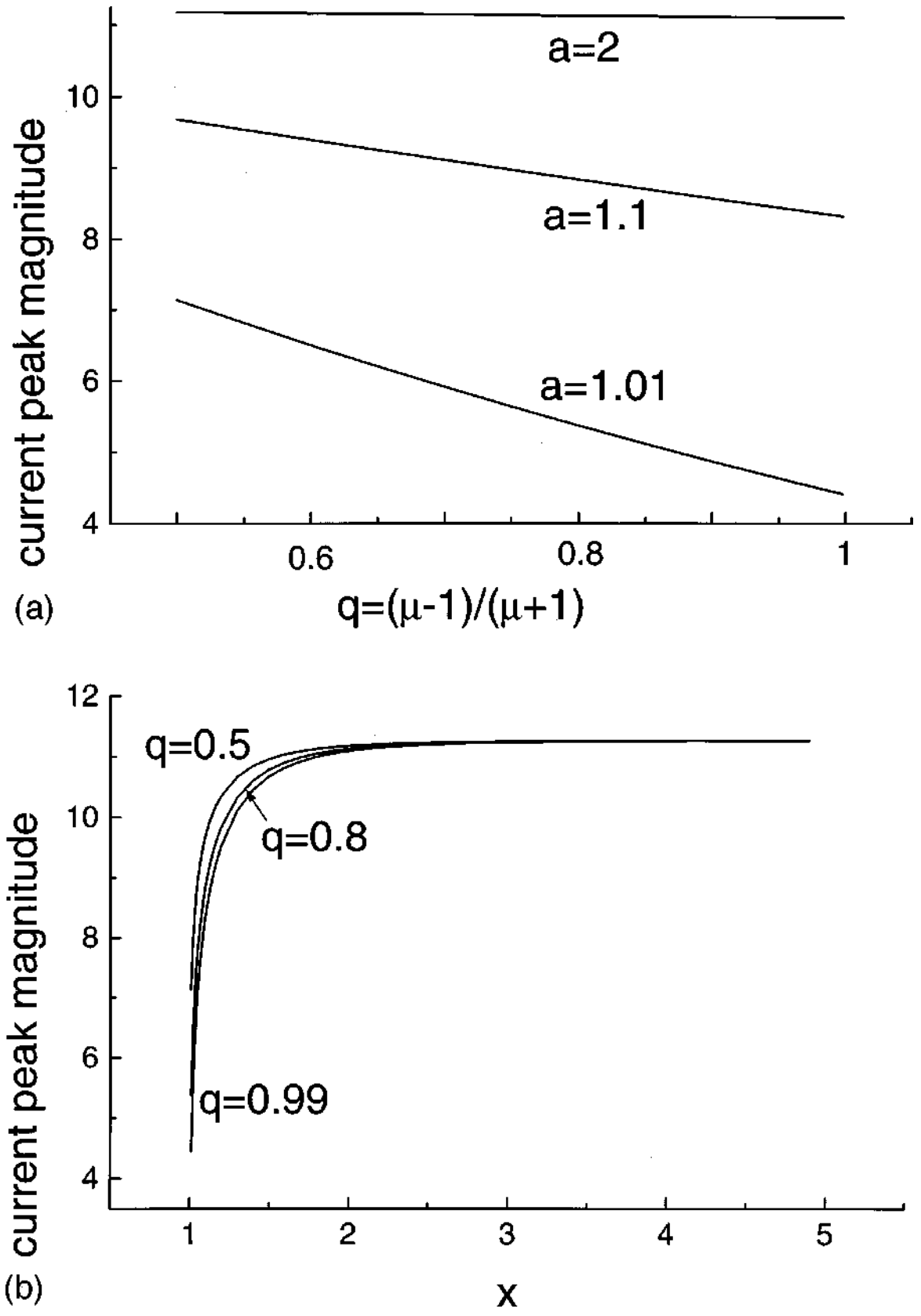

FIG. 6. (a) Dependence of the current-peak magnitude taken at the point $x=0.9996$ of a current sheet inside a cylindrical magnetic cavity versus the image current strength $q$. The total current $I=1$. Different curves correspond to cavity radii $a=2$, 1.1, and 1.01. (b) The same current-peak magnitude versus distance $a$ for different $q=0.5$, 0.8, and 0.99 ($\mu=3$, 9, and 200, respectively).

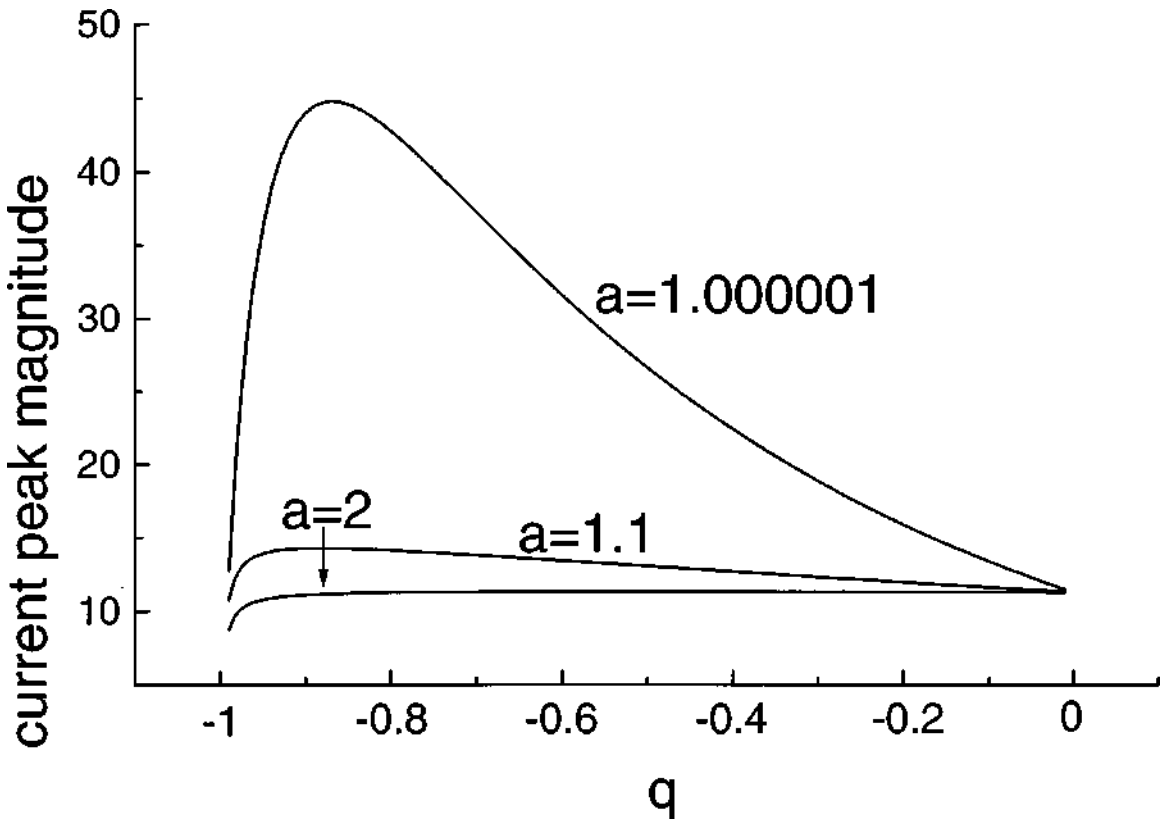

FIG. 7. Dependence of the current edge peaks on the negative parameter $q=(\mu_2-\mu_1)/(\mu_2+\mu_1)$ in the case of vacuum ($\mu_2=1$) surrounding a magnetic cylinder ($\mu_1>1$). Different curves correspond to different cylinder radii $a=1.000\,001$, 1.1, and 2.

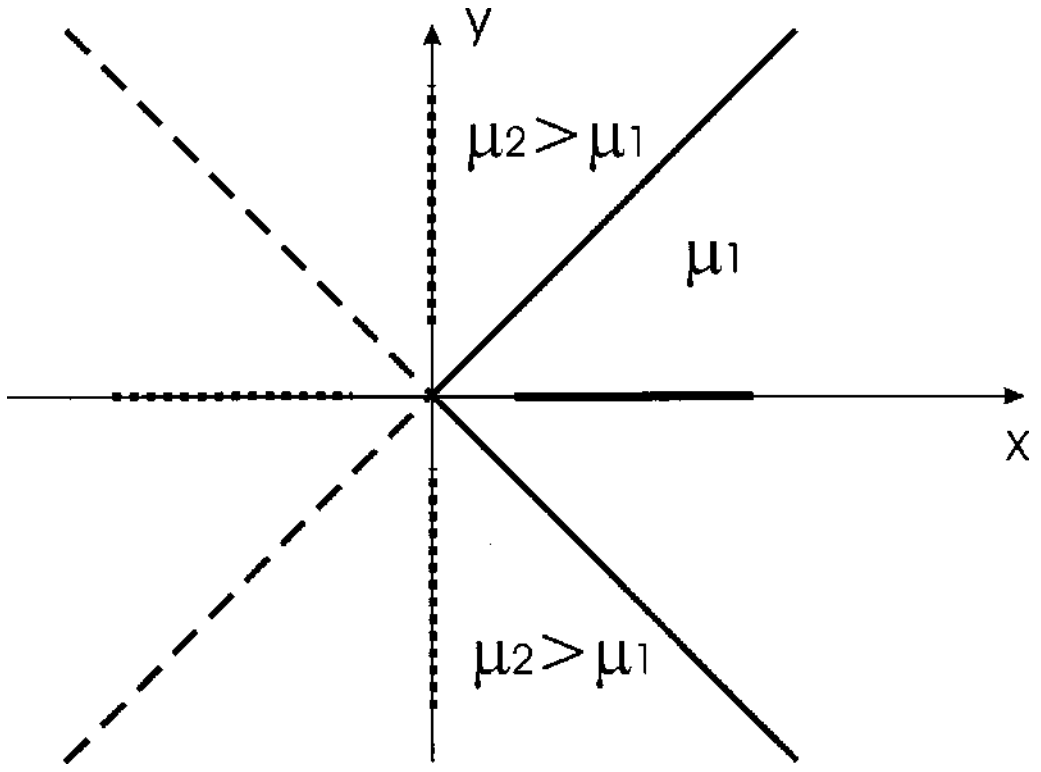

FIG. 8. Current sheet inside a right-angle wedge-shaped magnet cavity and its three images shown with dotted lines.

## III. MAGNETIC SCREENING OF THE MEISSNER STATE FOR MORE COMPLICATED GEOMETRIES WITH $\mu \gg 1$

The study of current redistributions under the effect of magnetic screens of a high but finite permeability $\mu$ carried out in Sec. II shows no crucial changes at $q \rightarrow 1$ ($\mu \rightarrow \infty$). This allows one to concentrate on the role of geometry and to assume for simplicity $q=1$. Under this assumption, we study in this section some definite geometries of magnetic surroundings which provide a strong suppression of current peaks at strip edges and thus an effective protection of the Meissner state.

In the limit $\mu \rightarrow \infty$ corresponding to $\varepsilon \rightarrow 0$ in analogous electrostatic problem, magnetic field lines seem to behave exactly as lines of the electrostatic field near the surface of a conductor. Indeed, making use of the formula (11) one can easily check that the tangential component of the field at the boundary surface vanishes as $1/\mu_2$ for $\mu_2 \rightarrow \infty$. This gives us the possibility to use the tools elaborated for the calculation of electrostatic fields around conductors, such as the method of images (partly used in Sec. II) or of conformal mapping. The analogy has though a restricted usage that will be discussed below.

### A. Current sheet in an open wedge-shaped cavity

We first study the simple case of the current sheet parallel to axis of the wedge-shaped cavity inside a soft magnetic medium of permeability $\mu_2 \gg 1$ (Fig. 8). We suppose that, in this case, the analogy with electrostatic fields around a metal is valid and that the field lines are perpendicular to the surface of the magnet from the side of cavity filled with material with $\mu_1 \ll \mu_2$.

If the angle between the intersecting planes in Fig. 8 is a rational fraction of $\pi$ the number of images of a straight current necessary to construct a solution satisfying Eq. (8)

and obeying the boundary conditions (9) is finite. We consider for simplicity the case of the cavity with a right angle between the planes. The solution may be presented by the superposition of contributions of the sheet itself and its three images as is shown in Fig. 8.

The corresponding vector potential in the cavity reads

$$A(x,y)=-\frac{\mu_1}{4\pi}\int_{a-W/2}^{a+W/2}du\,J(u)\{\ln[y^2+(x-u)^2]+\ln[y^2+(x+u)^2]+\ln[x^2+(y+u)^2]+\ln[x^2+(y-u)^2]\},\tag{25}$$

where the origin is shifted for convenience to the vertex of the cavity and $a$ denotes the distance between the vertex and the sheet center. The magnetic field in the cavity,

$$\mathbf{H}(x,y)=\frac{1}{2\pi}\int_{a-W/2}^{a+W/2}du\,J(u)\left[\frac{(-y,x-u)}{(x-u)^2+y^2}+\frac{(-y,x+u)}{(x+u)^2+y^2}+\frac{(u-y,x)}{x^2+(y-u)^2}+\frac{(-u-y,x)}{x^2+(y+u)^2}\right],\tag{26}$$

exhibits a zero tangential component at the boundaries $y=\pm x$ as was assumed.

The condition for the Meissner state formulates now as

$$H(x)=\frac{1}{2\pi}\int_{a-W/2}^{a+W/2}du\,J(u)\left(\frac{1}{x-u}+\frac{1}{x+u}+\frac{2x}{x^2+u^2}\right)=\frac{2x^3}{\pi}\int_{a-W/2}^{a+W/2}\frac{du\,J(u)}{x^4-u^4}=0.\tag{27}$$

Substituting a new variable $v=u^4$ and a new unknown function $\phi(v)=J(u)/4u^3$ into Eq. (27) one can reduce it to the one similar to Eq. (6) and finally find a solution:

$$J(x)=\frac{4Ix^3}{\pi\sqrt{[(a+W/2)^4-x^4][x^4-(a-W/2)^4]}}\tag{28}$$

satisfying the normalization (2).

One can note a much stronger suppression of the current peak at the sheet edge close to vertex of the cavity as compared to the flat surface of magnetic half space [see Eq. (14)]. For the sheet directly contacting the magnet ($a\to W/2$) one obtains a reduction of the sheet current distribution to

$$J(x)=\frac{4Ix}{\pi\sqrt{W^4-x^4}}.\tag{29}$$

The most remarkable feature of the above solution is the vanishing of the sheet current at the edge contacting the magnet. An analogous result was obtained in Ref. 2 for the current redistribution in the open magnetic cavity of the complicated form with wedgelike edges. Essentially, the current near the sheet edge behaves as $x^{1/p-1}$ where the wedge angle is $p\pi$. Practically, the complete vanishing of the edge current cannot be obtained because in any experiment a distance $a-W/2$ remains $>0$ and because the effect crucially depends on $(a-W/2)$ (see Sec. II B). Nevertheless, the wedge-shaped edge of the cavity provides a much more effective suppression of the current peaks at the edges and leads to a current redistribution to the center of the sheet as will be shown below.

### B. Conformal mapping of the screening cavity

The method of conformal mapping is very helpful in the solution of two-dimensional problems in electrostatics of conductors.[14,15] Its validity is based on two grounds: First, the conformal mapping keeps the form of the two-dimensional Laplace Eq. (8); second, it keeps the local angles between the curves in the $(x,y)$ plane after the transformation. Thus if field lines are perpendicular to a surface (of a conductor) they remain perpendicular to a transformed surface after the mapping. In other words, the latter conserves the boundary condition of the Neumann problem $\partial A/\partial n=0$.[24]

In the limit $\mu\gg1$, the conformal mapping may be used for finding field and current configurations in magnetostatics too but with precautions. The electrostatic field lines approach the metal surface always at right angles. The angles $\alpha_1$ and $\alpha_2$ which magnetic-field lines assume with the normal to the boundary surface between media 1 and 2 are connected by the refraction relation[14,15]

$$\frac{\tan\alpha_1}{\tan\alpha_2}=\frac{\mu_1}{\mu_2}.\tag{30}$$

As $\mu_2\to\infty$, this relation may be satisfied by two alternative solutions: $\alpha_1\to0$ while $0<\alpha_2<\pi/2$ which supports electrostatic analogy with the field pattern near the metal surface from the side of medium 1, and $\alpha_2\to\pi/2$ while $0<\alpha_1<\pi/2$ which gives another pattern of field lines. Both possibilities are realized for different geometries but, of course, only one option is valid in every case because of uniqueness theorem.[14,15]

It is easy to prove that the first option cannot be realized for a current $I$ flowing inside a closed cavity with medium 1 surrounded by medium 2 with $\mu_2\to\infty$. Indeed, if magnetic-field lines were perpendicular everywhere to the surface of the closed cavity, the contour integral along the surface $\oint\mathbf{H}d\mathbf{l}$ encircling the current sheet would be equal to zero instead of $I$ according to the Maxwell equation curl $\mathbf{H}=\mathbf{j}$. It is apparent that, for the same reason, the boundary condition $\partial A/\partial n=0$ cannot be applied to the internal surface of any closed cavity in a magnetic medium even at $\mu_2\to\infty$. In this case, the second alternative is realized and field lines are not perpendicular to the surface. That is why we will consider below only open configurations where the boundary condition $\alpha_1=0$ applies. One example of the open magnetic cavity with $\mu\to\infty$ was already considered in Sec. III A where $\alpha_1=0$ was supposed. One can easily prove that the solution (26) satisfies the Maxwell equation $\oint\mathbf{H}d\mathbf{l}=I$.

Let us introduce complex variables $s=x+iy$ and $w=\zeta+i\eta$ and some analytical function $w=f(s)$ carrying out the conformal mapping of the plane $(x,y)$ onto the plane $(\zeta,\eta)$. If a vector potential $A$ satisfies the equation $\Delta A=0$ around the current sheet with a condition $\partial A/\partial n=0$ at some boundaries then a vector-potential $\Omega(\zeta,\eta)=A[x(\zeta,\eta),y(\zeta,\eta)]$

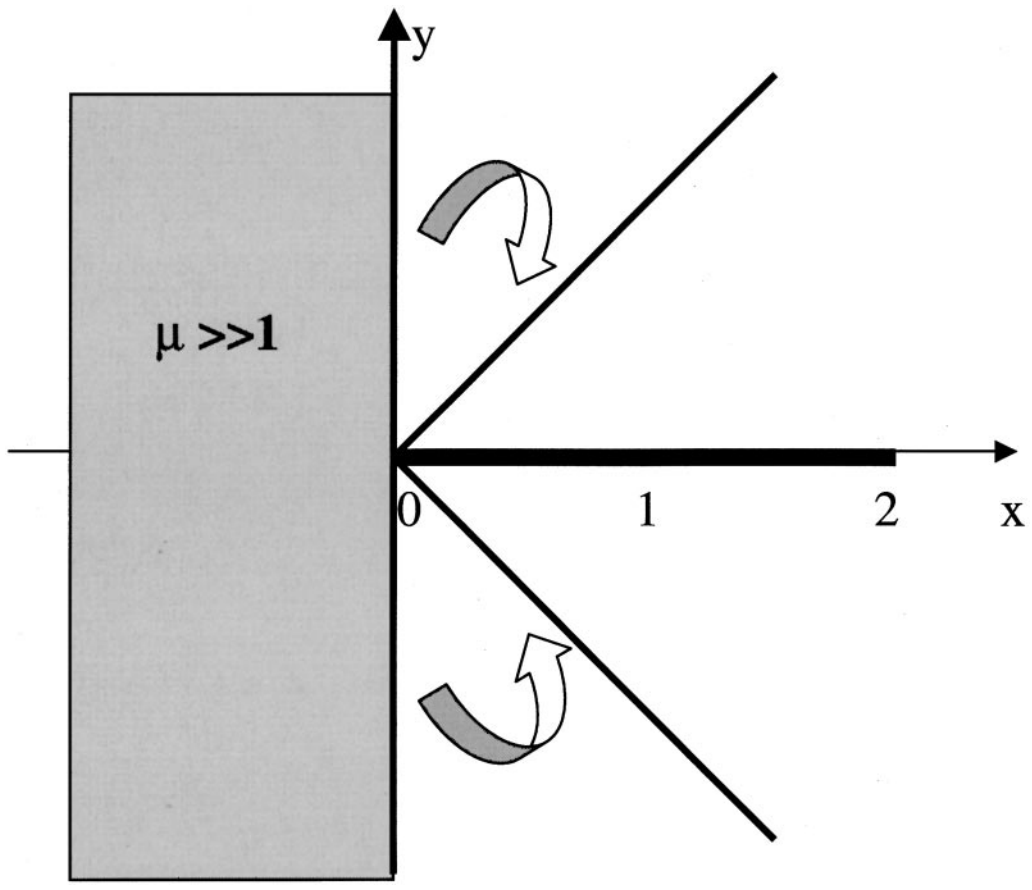

FIG. 9. Scheme of the conformal mapping of the magnetic half space boundary $x=0$ to the surface of a right-angle wedge-shaped cavity.

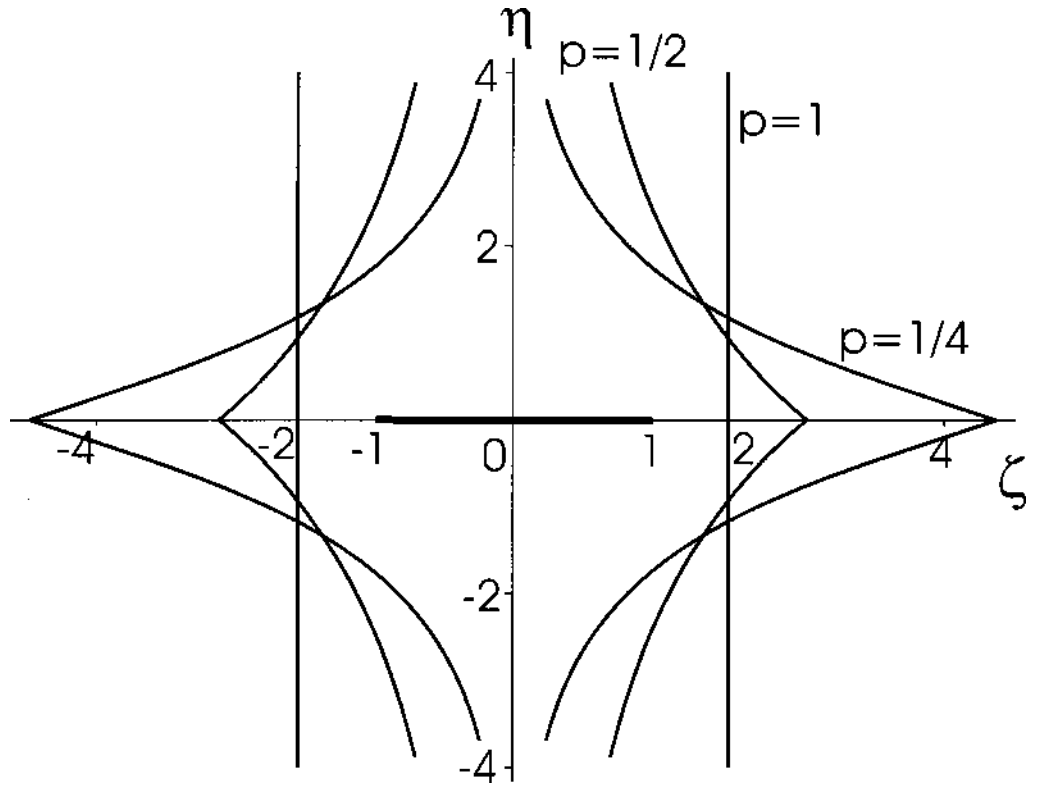

FIG. 10. Scheme of the conformal mapping of the two surfaces of magnetic half spaces $x\geqslant a$ and $x\leqslant -a$ onto the surface of the convex cavity keeping the position of the current sheet at $(-1,1)$ for $a=2$ and different values of parameter $p=1/2$ and 1/4.

satisfies the equation $\Delta\Omega=0$ in the new frame $(\zeta,\eta)$ and condition $\partial\Omega/\partial n=0$ at the transformed boundaries as well.[24] Thanks to conservation of local angles between the curves the field lines parallel to the current sheet in $(x,y)$ plane remain parallel to a transformed sheet also in $(\zeta,\eta)$ plane. This means that the conformal mapping conserves also a Meissner state of the sheet. Thus any conformal mapping of the current sheet being in the Meissner state which keeps the position of the sheet will give us another possible pattern of the Meissner state of the sheet for another geometry of the surrounding.

Let us denote the magnetic field in new coordinates by $\tilde{\mathbf{H}}=(\partial\Omega/\partial\eta,-\partial\Omega/\partial\zeta)$. Then the new Meissner current distribution $\tilde{J}$ may be found as a jump of the parallel field component at the sheet [see Eq. (4)],

$$\tilde{J}(\zeta)=\tilde{H}_\zeta(\zeta,\eta)\Big|_{\eta=+0}^{\eta=-0}=\frac{\partial y}{\partial\eta}\Big|_{\eta=0}J[x(\zeta,\eta=0)]. \tag{31}$$

Let us consider first a case of transformation of a flat boundary of the bulk magnetic medium with $\mu\gg 1$ to the wedge-shaped configuration treated in the previous section in terms of images (Fig. 8). Let a magnet fill the left half space $x\leqslant 0$ and the current sheet which occupies the position $0\leqslant x\leqslant W$ directly contact the magnet. According to Sec. II B, the current distribution delivering the Meissner state ($H_y=0$ at the sheet) is given by the formula

$$J(x)=\frac{2I}{\pi\sqrt{W^2-x^2}},\quad 0\leqslant x\leqslant W, \tag{32}$$

obtained from Eq. (15) by symmetrical reflection and shift of the origin.

As is well known,[24] the transformation of the half plane into the quarter of plane shown in Fig. 9 is carried out by a square-root function. Scaled to keep the sheet position on the horizontal axis it is described by a function $f(s)=\sqrt{Ws}$. Then, finding the derivative $\partial y/\partial\eta$ from the relation $x+iy=(\zeta+i\eta)^2/W$ and substituting it into Eq. (31) one finds a Meissner current in the new geometry

$$\tilde{J}(\zeta)=\zeta J[x(\zeta,\eta=0)]=\frac{4I\zeta}{\pi\sqrt{W^4-\zeta^4}},\quad 0\leqslant\zeta\leqslant W, \tag{33}$$

which reproduces the solution (29).

For the general case of a wedge-shaped magnetic cavity with a wedge angle $\alpha=p\pi$, $0<p<1$ one needs an analytical function $f(s)=W(s/W)^p$. This gives a derivative $\partial y/\partial\eta\rightarrow p^{-1}(\zeta/W)^{1/p-1}$ at $\eta=0$ which results in the Meissner current [see Eq. (31)]

$$\tilde{J}(\zeta)=\frac{2I}{p\pi W}\frac{(\zeta/W)^{1/p-1}}{\sqrt{1-(\zeta/W)^{2/p}}},\quad 0\leqslant\zeta\leqslant W. \tag{34}$$

This shows that in a (hardly experimentally accessible) case of a sharp cavity with $\alpha\ll\pi(p\ll 1)$ the current would be well suppressed in the extended region near the edge.

The practically important conformal mapping of two flat surfaces onto the complicated concave cavity was considered in Ref. 2 only for the case of direct contact magnet-superconductor. Here we generalize this transformation to study the effect of the cavity size and distance to the strip on the current redistribution. We start from the current sheet between the two magnetic half spaces possessing $\mu\rightarrow\infty$ with flat boundaries located at $x=\pm a$. As was shown in Ref. 2, the Meissner current distribution for this geometry is given by the formula

$$J(x)=\frac{I}{a}\frac{\cos(\pi x/2a)}{\sqrt{2\cos(\pi x/a)-2\cos(\pi W/2a)}}. \tag{35}$$

Let us perform a mapping of the plane $s$ using an analytical function

$$f(s)=c_a[(1+s/a)^p-(1-s/a)^p], \tag{36}$$

where $c_a=(W/2)[(1+W/2a)^p-(1-W/2a)^p]^{-1}$. This mapping moves the points $s=\pm W/2$ to $w=\pm W/2$ and thus conserves the position of the current sheet (Fig. 10). The edges of the cavity $s=\pm a$ are moved to the points $w=\pm 2^p c_a$. The mapping of the form (36) allows one to control the angles $\alpha=p\pi$ at the edges of the cavity and thus the corresponding current distribution. We consider below for simplicity only the case $p=1/2$.

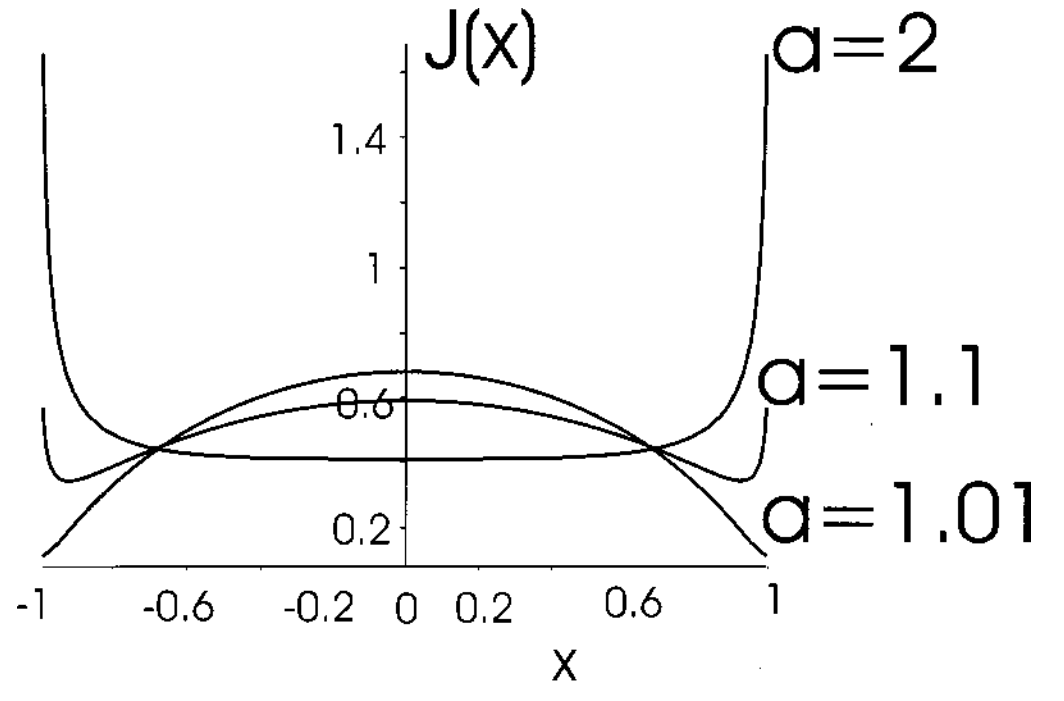

FIG. 11. Transport current distributions over the flux-free superconducting sheet in the convex magnetic cavity with $p=1/2$ at different distances between the magnet and the sheet center $a=2$, 1.1, 1.01.

Making use of the formula (31) we find from Eq. (36)

$$\frac{\partial y}{\partial \eta}\Big|_{\eta=0}=\frac{a}{c_a}\frac{1-\zeta^2/2c_a^2}{\sqrt{1-\zeta^2/4c_a^2}}, \tag{37}$$

which gives for the corresponding Meissner current

$$\tilde{J}(\zeta)=\frac{I}{c_a}\frac{1-\zeta^2/2c_a^2}{\sqrt{1-\zeta^2/4c_a^2}} \times\frac{\cos\left(\frac{\pi\zeta}{2c_a}\sqrt{1-\zeta^2/4c_a^2}\right)}{\sqrt{2\cos\left(\frac{\pi\zeta}{c_a}\sqrt{1-\zeta^2/4c_a^2}\right)-2\cos(\pi W/2a)}}. \tag{38}$$

The patterns of the above current distribution for various cavity transverse sizes are shown in Fig. 11. One can follow the evolution of the current pattern from a concave distribution with sharp edge peaks typical of an isolated strip in the Meissner state to the convex distribution with suppressed peaks at the edges upon decreasing the transverse cavity size.

The magnetic field in this geometry remains perpendicular to the cavity surface in no contradiction with the Maxwell equation. To check this, consider the simplest case $a=W/2$. Let us carry out the contour integration of the field along the path following the cavity surface and closed across the narrow "bottlenecks" of the cavity of the width $\simeq 1/\eta$ at $\eta \gg 1$ as is shown in Fig. 12. The contribution of the surface part equals zero since the field is normal to the surface. The magnetic field at the symmetry axis of the cavity ($\zeta=0$) equals asymptotically $\tilde{H}_\zeta \simeq I\eta/2$ at $\eta \gg 1$, i.e., grows to infinity. Then, integrating along the infinitesimal bottlenecks one finds $\oint \tilde{\mathbf{H}}d\mathbf{l}=I$.

The comparison of the open magnetic surroundings shown in Figs. 8 and 10 with the closed one (Fig. 4) and of the corresponding current distributions presented in Figs. 2, 5, and 11 allows one to conclude that field lines perpendicular to the magnet surface (which is valid for open cavities only) may be more effectively controlled by a form of the cavity. In case of the closed cavity, one can control the field (and hence current) distribution to a much lesser extent.

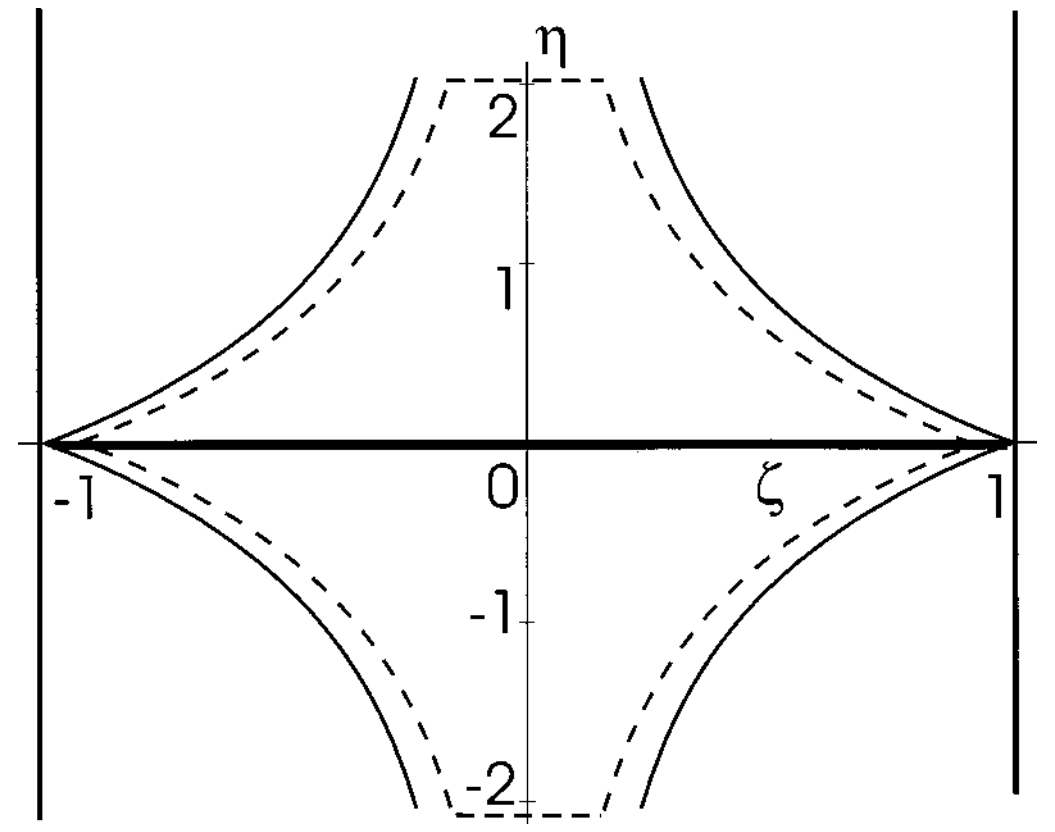

FIG. 12. Scheme of the conformal mapping of the two surfaces of magnetic half spaces $x \geq 1$ and $x \leq -1$ onto the surface of the convex cavity with a parameter $p=1/2$. The contour of integration for the testing a solution is shown with a dashed line.

Let us estimate the maximal transport current which may flow in the magnetically screened flux-free superconducting strips in various magnetic surroundings. For the strip placed between the flat magnets as in Fig. 10 ($p=1$), the peak value of the sheet current at the edge point $x_\Gamma=W/2-\delta=0.9996 \cdot W/2$ follows from Eq. (35) and equals

$$J_\Gamma=\frac{I}{\sqrt{4\pi a\delta\tan(\pi W/4a)}}. \tag{39}$$

Equating this with the sheet current of the geometrical barrier $J_b$ one finds the maximum averaged current density which can flow in the Meissner state as a function of distance $a$:

$$j_M=\frac{I}{Wd}=\frac{2J_b}{Wd}\sqrt{4\pi a\delta\tan(\pi W/4a)}, \tag{40}$$

where $\delta < a-W/2 \ll W/2$. One sees an increase of $j_M$ for $a \rightarrow W/2$. Assuming a least possible gap between the magnet and the sheet of about 0.1 mm one finds, for the above considered YBCO film of width $W=5$ mm and thickness $d=50$ nm (see Sec. II A), a value $j_M=3.4\times 10^6$ A/cm$^2$.

The screening by the convex cavity (Figs. 10 and 11) enlarges the bearable transport current in the flux-free state more effectively. The current peak value in this configuration follows from Eq. (38) at $\zeta_\Gamma=W/2-\delta$ and equals

$$J_\Gamma=I\sqrt{\frac{1-W^2/8c_a^2}{\pi W\delta\tan\left(\frac{\pi}{2c_a}\sqrt{1-W^2/16c_a^2}\right)}}\simeq\frac{4I}{W^2}\frac{(r-W/2)^{3/2}}{\sqrt{\delta}}, \tag{41}$$

where $r=c_a\sqrt{2}$ is a transverse size of the cavity. Upon the equating $J_\Gamma=J_b$ this gives the maximum average current density

$$j_M=\frac{J_\Gamma}{d}\frac{W\sqrt{\delta}}{4(r-W/2)^{3/2}}, \quad \delta < r-\frac{W}{2} \ll \frac{W}{2}, \tag{42}$$

which results in $j_M = 2.9\times 10^7$ A/cm$^2$ by $2r/W = 1.04$ for the same YBCO film. This value is about the depairing limit for this material at $T = 77$ K which switches on a process of direct destruction of superconducting pairs[25] in the strip center where the current is maximum. To avoid this the cavities of more complicated shapes should be considered which allows one to suppress effectively the current peaks at the sheet edges and provide a homogeneous filling of the sheet with the transport current.

## IV. PARTLY FLUX-FILLED HARD SUPERCONDUCTOR STRIPS IN A FLAT MAGNETIC ENVIRONMENT

In previous sections we considered the Meissner flux-free state in the case of a pronounced edge barrier with $J_b > J_c$ valid in single crystals of BSCCO (Refs. 5 and 6) and YBCO.[21] In the following sections, we consider the opposite case of the negligible edge barrier and characteristic current $J_b < J_c$ which is typical of not too thin films of high-$T_c$ materials.

A conventional tool for the description of the electrodynamics of hard superconductors is the critical-state model originally applied to long bulk superconductors in a parallel field.[1,26] According to this model, magnetic-flux lines of an external field or generated by a transport current enter the sample and move until the Lorentz force density $|\mathbf{j}\times\mathbf{B}|$ exerted upon them by the local current density $\mathbf{j}$ reaches the pinning-mediated critical value $j_c(B)\cdot B$, where $\mathbf{B}$ is a magnetic-flux density. Then the flux lines stop to move and form flux profiles determined by the equation $j = |\nabla\times\mathbf{H}| = j_c(B)$, where $\mathbf{H} = \mathbf{B}/\mu_0$ is a local magnetic field.

In the simplest Bean model[26] of the critical state which we use in the following, $j_c$ is supposed to be field independent. Then, in the case of full flux penetration, $|j| = j_c$ all over the long bulk sample. Since in the case of a long slab or cylinder parallel to the field the latter has only one component, the current and field distributions in the partly flux-filled state are related to each other locally. The current density and the flux density are simultaneously equal to zero in the flux-free region of the specimen.[1,26] In the case of a magnetic field applied transverse to a flat superconducting sample the field lines are strongly curved, the current-field relation becomes nonlocal and the above simple picture is no longer valid.

Current and field distributions in flat samples, especially in the partly flux-filled state, turned out to look very different from those in bulk samples.[10–13,27–32] The most remarkable difference is that shielding currents flow along the entire width of the sample including the flux-free region whereby the field component perpendicular to the surface is equal to zero.

Direct magnetooptical observations of the flux distributions in wide superconducting strips subjected to an external field[33] or carrying a transport current[34] and current distributions extracted from such data[35] are in good agreement with theoretical predictions.[12,13] Before we begin with the study of flux distributions in magnetically shielded hard superconductors it is necessary to resume the main features of the partly flux-filled state of an isolated superconductor strip.

### A. Partly flux-filled state of an isolated strip

Consider now a current of magnitude $I$ such that the sheet current peak at the edge $J_\Gamma > J_c = j_c d$ and magnetic flux partly penetrates the strip. According to the Bean hypothesis of force balance,[26] the flux is distributed in a way that the current is constant in the penetrated region. The rest of the sample remains in the flux-free state so that

$$H(x) = 0, \quad |x| < b < W/2,$$

$$J = J_c, \quad b < |x| < W/2, \tag{43}$$

where the width $2b$ of the flux-free region is determined by the value of the total current $I$.

Following the scheme suggested by Norris[10] the solution to this problem may be found as follows. First, one should find a current distribution arising in a flux-free region $|x| \leqslant b$ in response to a current of unit strength flowing along the line $x = x_0 > b$. Then the state satisfying requirements (43) is obtained by superposition of responses to all current filaments of strength $J_c\cdot dx_0$ in a constant-$J_c$ region $b < |x_0| < W/2$.

Since in the following we consider only bilaterally symmetrical problems we are interested in the even current distributions. Then we may seek an equation for a current shielding the region $|x| \leqslant b$ from the field induced by two test unit currents located at $x = \pm x_0$ by substitution of

$$J = J(u)\cdot\theta(b - |x|) + \delta(u - x_0) + \delta(u + x_0) \tag{44}$$

into Eq. (6). One finds then an inhomogeneous singular integral equation of the form

$$\int_{-b}^{b} \frac{du\, J(u)}{u - x} = g(x) = \left(\frac{1}{x - x_0} + \frac{1}{x + x_0}\right), \quad |x| < b. \tag{45}$$

The general solution to this equation with arbitrary right-hand side (r.h.s.) $g(x)$ reads[22]

$$J(x) = \frac{1}{\pi^2\sqrt{b^2 - x^2}} \int_{-b}^{b} \frac{du\,\sqrt{b^2 - u^2}}{x - u} g(u) + \frac{N}{\pi\sqrt{b^2 - x^2}} \tag{46}$$

with some arbitrary constant $N$.

Substituting the right-hand side from Eq. (45) into Eq. (46) one can reproduce the result found by Norris by means of conformal mapping:

$$J(x, x_0) = -\frac{2x_0\sqrt{x_0^2 - b^2}}{\pi(x_0^2 - x^2)\sqrt{b^2 - x^2}}. \tag{47}$$

Integrating Eq. (47) over the test current position $x_0$ in the constant-$J_c$ region $b < x_0 < W/2$ and adding the term $\sim (b^2 - x^2)^{-1/2}$ originating from a solution (46) to homogeneous Eq. (45) to remove the unphysical divergencies at $x = \pm b$ one finds finally a current distribution in a partly flux-filled state:[10]

$$J(x)=\frac{2J_c}{\pi}\arctan\sqrt{\frac{(W/2)^2-b^2}{b^2-x^2}},\quad |x|<b<W/2,$$
$$J_c,\quad b<|x|<W/2. \tag{48}$$

The total current may be found by integration of the above sheet current over the strip width. This gives

$$I=I_c\sqrt{1-(2b/W)^2}, \tag{49}$$

where $I_c=J_c\cdot W$ is the maximum nondissipative current in the critical state achieved by a saturation of the strip with magnetic flux. Inversely, one can find from Eq. (49) the half-width of the flux-free zone $b=(W/2)\sqrt{1-(I/I_c)^2}$.

The corresponding distribution of the perpendicular component of the field over the strip may be found by substituting the above current into Eq. (5),[12,13]

$$H(x)=0,\quad |x|<b,$$
$$\frac{H_c x}{|x|}\operatorname{arctanh}\sqrt{\frac{x^2-b^2}{(W/2)^2-b^2}},\quad b<|x|<\frac{W}{2},$$
$$\frac{H_c x}{|x|}\operatorname{arctanh}\sqrt{\frac{(W/2)^2-b^2}{x^2-b^2}},\quad |x|>\frac{W}{2} \tag{50}$$

where $H_c=J_c/\pi$ denotes a characteristic field.

The above field and current distributions apply to the virgin state where the transport current $I$ is increased from zero. More complicated history-dependent distributions are considered in details in Refs. 12 and 13.

### B. Superconductor strip between flat magnets: Parallel configuration

In this section we study the current and field distributions of a current-carrying superconductor strip partly penetrated by magnetic flux and placed between two half spaces occupied by soft magnets. The medium directly contacting the strip is characterized by magnetic permeability $\mu_1$, the magnets by a permeability $\mu_2$. This two-dimensional magnetostatic problem may be as well conveniently treated by means of Eq. (8) for the vector potential $\mathbf{A}$ with boundary conditions (9). General expressions will be written down for arbitrary $\mu_1$ and $\mu_2$, but explicit solutions for current distributions will be found only for the case $\mu_2 \gg \mu_1$.

We begin with the configuration when the boundaries of magnets $y=\pm a$ are parallel to the strip and the magnets occupy the space at $\pm y \geqslant a$. The magnetic field $\mathbf{H}$ and potential $A$ satisfying the proper boundary conditions may be constructed in this case by means of the method of images[14,15] as a superposition of the contributions from the strip itself and its images at the ''mirrors'' at $y=\pm a$. The corresponding set of images represents a series of parallel equidistant strips centered at $y=2an$, $n=\pm1,\pm2,\ldots$ each carrying the same current in the same direction (see Fig. 13).

Then the vector potential in between the magnets reads

$$A_1=-\frac{\mu_1}{4\pi}\sum_n q^{|n|}\int_{-W/2}^{W/2}du\,J(u)\ln[(y-2an)^2+(u-x)^2], \tag{51}$$

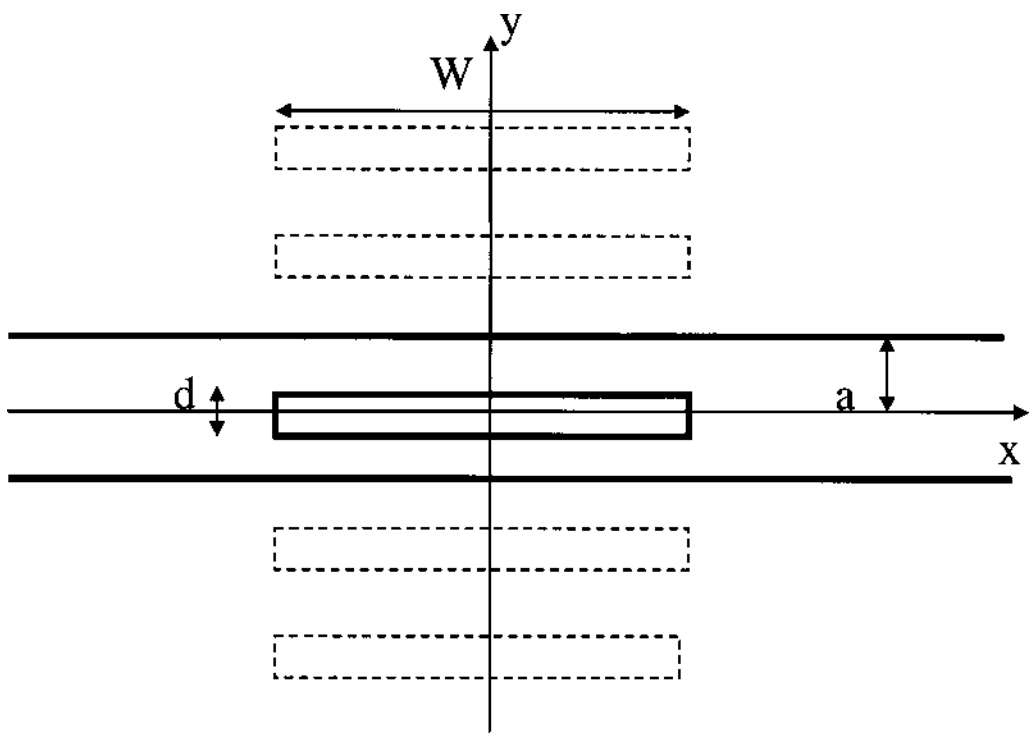

FIG. 13. (a) The current-carrying sheet between two magnet half spaces with boundaries parallel to the sheet. All surfaces are shown with solid lines and images of the sheet at the magnetic mirrors are shown with dashed lines.

where the summation extends over all integers, and inside the magnets

$$A_2=-\frac{\mu_1\mu_2}{2\pi(\mu_1+\mu_2)}\sum_{n=0}^{\infty}q^n\int_{-W/2}^{W/2}du\,J(u)\ln[(y\pm2an)^2+(u-x)^2],\quad \pm y\geqslant a. \tag{52}$$

The magnetic field $\mathbf{H}$ is distributed in between the magnets according to the expression

$$\mathbf{H}(x,y)=\frac{1}{2\pi}\sum_n q^{|n|}\int_{-W/2}^{W/2}du\,J(u)\frac{(2an-y,x-u)}{(x-u)^2+(y-2an)^2}. \tag{53}$$

For the case $\mu_2 \gg \mu_1$ one can calculate the above series at $q\to1$,[36] and find

$$\mathbf{H}(x,y)=\frac{1}{4a}\int_{-W/2}^{W/2}du\,J(u)\frac{\left[-\sin\frac{\pi}{a}y,\sinh\frac{\pi}{a}(x-u)\right]}{\cosh\frac{\pi}{a}(x-u)-\cos\frac{\pi}{a}y}. \tag{54}$$

For the perpendicular field component at the sheet surface one obtains

$$H(x)=\frac{1}{4a}\int_{-W/2}^{W/2}du\,J(u)\coth\left(\frac{\pi}{2a}(x-u)\right). \tag{55}$$

All the above formulas in this section were derived so far for an arbitrary current distribution $J(x)$. Now we apply these to a description of the partly flux-filled state of the strip. For this aim, we substitute the field expression (55) into Eqs. (43). The resulting problem is very close to the problem of field and current distributions over a stack of superconducting films considered by Mawatari.[32] In fact, the set of images in the limit of infinite permeability $\mu\to\infty$($q\to1$) is, in the region $|y|\leqslant a$, physically identical to the stack of real films. The only difference from results of Mawatari consists in the fact that in his work a stack of films was considered exposed to an external field in the absence of a transport current while we study the alternative case of the transport current applied to a strip in the absence of an external field.

As well as in Ref. 32 one can introduce new variables $\tau = \tanh \pi u/2a$ and $t = \tanh \pi x/2a$, which vary within the interval $(-l,l)$ with $l = \tanh \pi W/4a$, and parameter $p = \tanh \pi b/2a$, and define a new unknown function $\phi(\tau) \equiv J(u)$. Substituting the new variables into Eq. (55) and seeking for the even solution $J(u) = J(-u)$, which is followed by $\phi(\tau) = \phi(-\tau)$, one finds conditions of the partly flux-filled state (43) for the new variables

$$H(x) = \frac{1}{2\pi}\int_{-l}^{l} \frac{d\tau \phi(\tau)}{t-\tau} = 0, \quad |t| < p < l,$$

$$J = J_c, \quad p < |t| < l. \tag{56}$$

The identity of the above problem to that of the isolated strip [Eqs. (5) and (43)] allows one to write immediately the current distribution over the strip

$$J(x) = \frac{2J_c}{\pi}\arctan\sqrt{\frac{l^2-p^2}{p^2-\left(\tanh\frac{\pi}{2a}x\right)^2}}, \quad |x| < b,$$

$$J_c, \quad b < |x| < W/2 \tag{57}$$

and the field distribution over the strip

$$H(x) = 0, \quad |x| < b,$$

$$\frac{H_c x}{|x|}\mathrm{arctanh}\sqrt{\frac{\left(\tanh\frac{\pi}{2a}x\right)^2 - p^2}{l^2-p^2}}, \quad b < |x| < \frac{W}{2},$$

$$\frac{H_c x}{|x|}\mathrm{arctanh}\sqrt{\frac{l^2-p^2}{\left(\tanh\frac{\pi}{2a}x\right)^2 - p^2}}, \quad |x| > \frac{W}{2}. \tag{58}$$

The width of the flux-free region $2b$ is defined by the value of the total current. This may be found by integration of the above current distribution over the strip width

$$p = \sinh\left[\frac{\pi W}{4a}\left(1-\frac{I}{I_c}\right)\right] \times \sqrt{2\tanh\frac{\pi W}{4a}\left[\coth\frac{\pi W}{4a}\left(1-\frac{I}{I_c}\right) - \coth\frac{\pi W}{2a}\right]},$$

$$b = \frac{2a}{\pi}\mathrm{arctanh}(p). \tag{59}$$

An evolution of the current and field distributions with a decreasing distance to the magnets $a$ is shown in Fig. 14. Different curves present the same total current distributed over the strip for different distances $a$. At $a \gg W$ the distributions reproduce the current and field patterns of the isolated strip (48) and (50). The same result may easily be established analytically by considering $a \to \infty$ in Eqs. (57) and (58). At $a \simeq W$ the dip in the current distribution becomes deeper and narrower. At $a \ll W$ the current is almost completely expelled from the flux-free central part to the flux-filled regions along edges of the strip.

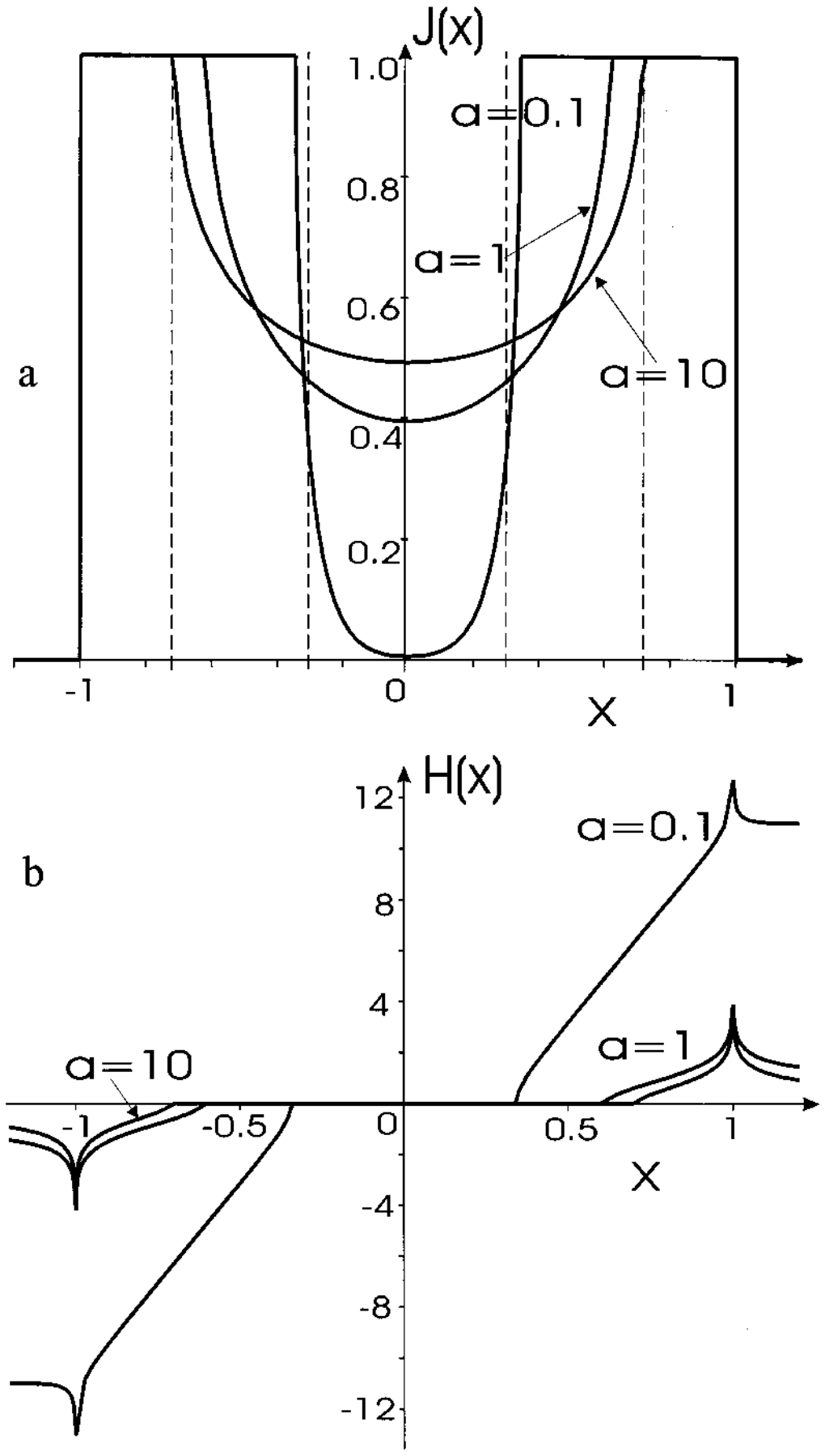

FIG. 14. Transport current (a) and perpendicular to the strip magnetic field (b) distributions over the sheet carrying the total current $I = 0.7 \cdot I_c$ are shown for various distances $a$ from the sheet to the magnets in parallel configuration. Vertical dashed lines at $x = \pm 0.714$ show the boundaries of flux-free region for $a = \infty$. Dashed lines at $x = \pm 0.3$ show the current profile at $a = 0$ which coincides with that of a slab of the same width.

The exhibited behavior is similar to the current distribution in a completely flux-free strip influenced by a magnetic surrounding studied in Ref. 2. By small $a$, the resulting current and field distributions approach the profiles typical for a long bulk slab.[1,26] This is not surprising since the set of strip images in our problem, as well as the stack of the films considered by Mawatari,[32] really form a slab of the width $W$ for $a \to 0$.

### C. Superconductor strip between flat magnets: Perpendicular configuration

Let us consider now the case of a transverse position of the strip with respect to the magnets as is shown in Fig. 15. The boundaries of the magnets assume now the locations $x = \pm a$ where $a > W/2$ holds for this geometry. The images of the strip lie now on the $x$ axis centered at $x = 2an$ with integer $n$. The solution to Eq. (8) in between the magnets now reads [even $J(u)$ is assumed]

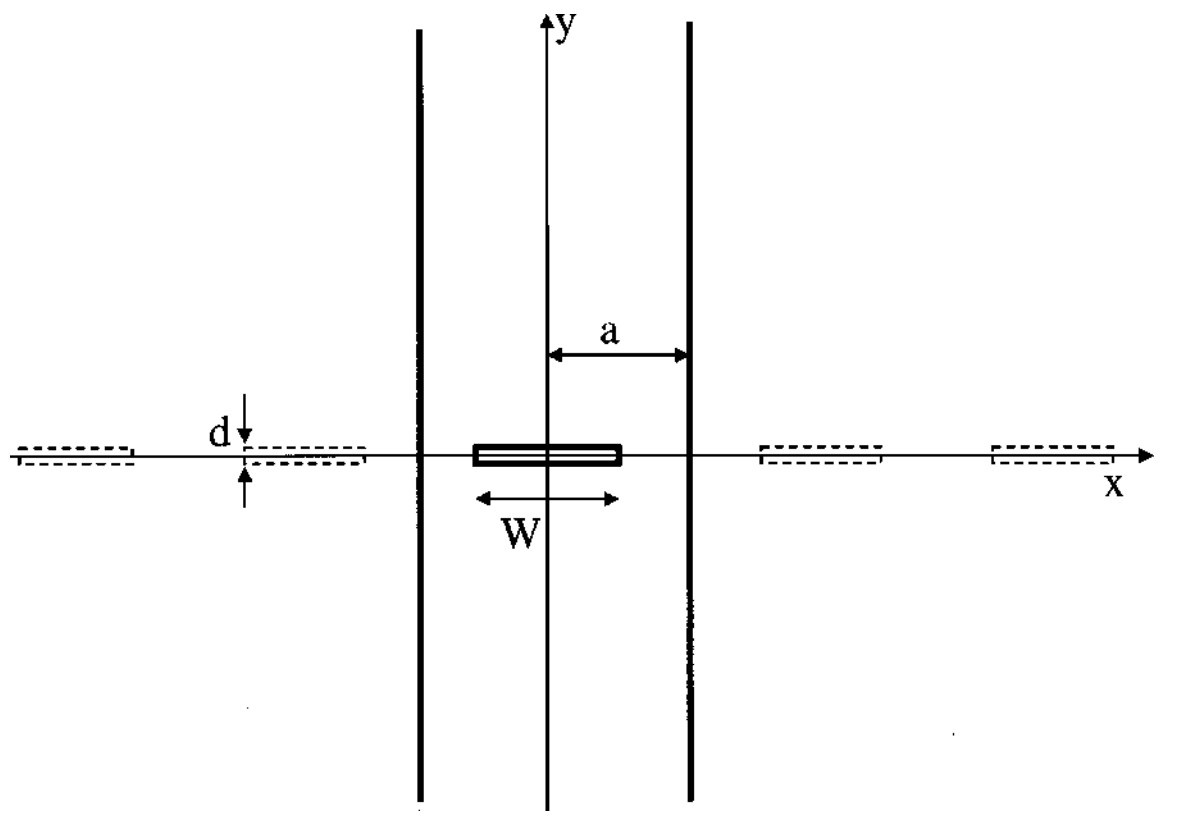

FIG. 15. The current-carrying sheet between two magnet half spaces with boundaries perpendicular to the sheet. All surfaces are shown with solid lines and images of the sheet at the magnetic mirrors are shown with dashed lines.

$$A_1=-\frac{\mu_1}{4\pi}\sum_n q^{|n|}\int_{-W/2}^{W/2}du\,J(u)\ln[y^2+(x-u-2an)^2], \tag{60}$$

and inside the magnets it is

$$A_2=-\frac{\mu_1\mu_2}{2\pi(\mu_1+\mu_2)}\sum_{n=0}^{\infty} q^n\int_{-W/2}^{W/2}du\,J(u)\times\ln[y^2+(x-u\pm 2an)^2],\quad \pm x\geqslant a. \tag{61}$$

Following from Eq. (60), the magnetic field in between the magnets now writes as

$$\mathbf{H}(x,y)=\frac{1}{2\pi}\sum_n q^{|n|}\int_{-W/2}^{W/2}du\,J(u)\frac{(-y,x-u-2an)}{(x-u-2an)^2+y^2}. \tag{62}$$

At $\mu_2\gg\mu_1$ one can put $q=1$ and find[36]

$$\mathbf{H}(x,y)=\frac{1}{4a}\int_{-W/2}^{W/2}du\,J(u)\frac{\left[-\sinh\frac{\pi}{a}y,\sin\frac{\pi}{a}(x-u)\right]}{\cosh\frac{\pi}{a}y-\cos\frac{\pi}{a}(x-u)}. \tag{63}$$

The perpendicular component of the field at the strip equals

$$H(x)=\frac{1}{4a}\int_{-W/2}^{W/2}du\,J(u)\cot\left(\frac{\pi}{2a}(x-u)\right). \tag{64}$$

The expressions (60)–(64) are again valid for an arbitrary current distribution. Now we try to find the current satisfying the conditions of the flux-filled state (43). This equation is similar to Eq. (55) and may be treated as in Mawatari's work[32] devoted to the field and current distributions over a horizontal array of superconductor films subjected to an external field. Substituting new variables and parameters $\tau=\tan\pi u/2a$, $t=\tan\pi x/2a$, $p=\tan\pi b/2a$, $l=\tan\pi W/4a$, and $\phi(\tau)\equiv J(u)$, one can once more reduce the conditions (43) exactly to Eqs. (56). This allows one to write a solution at once in the form

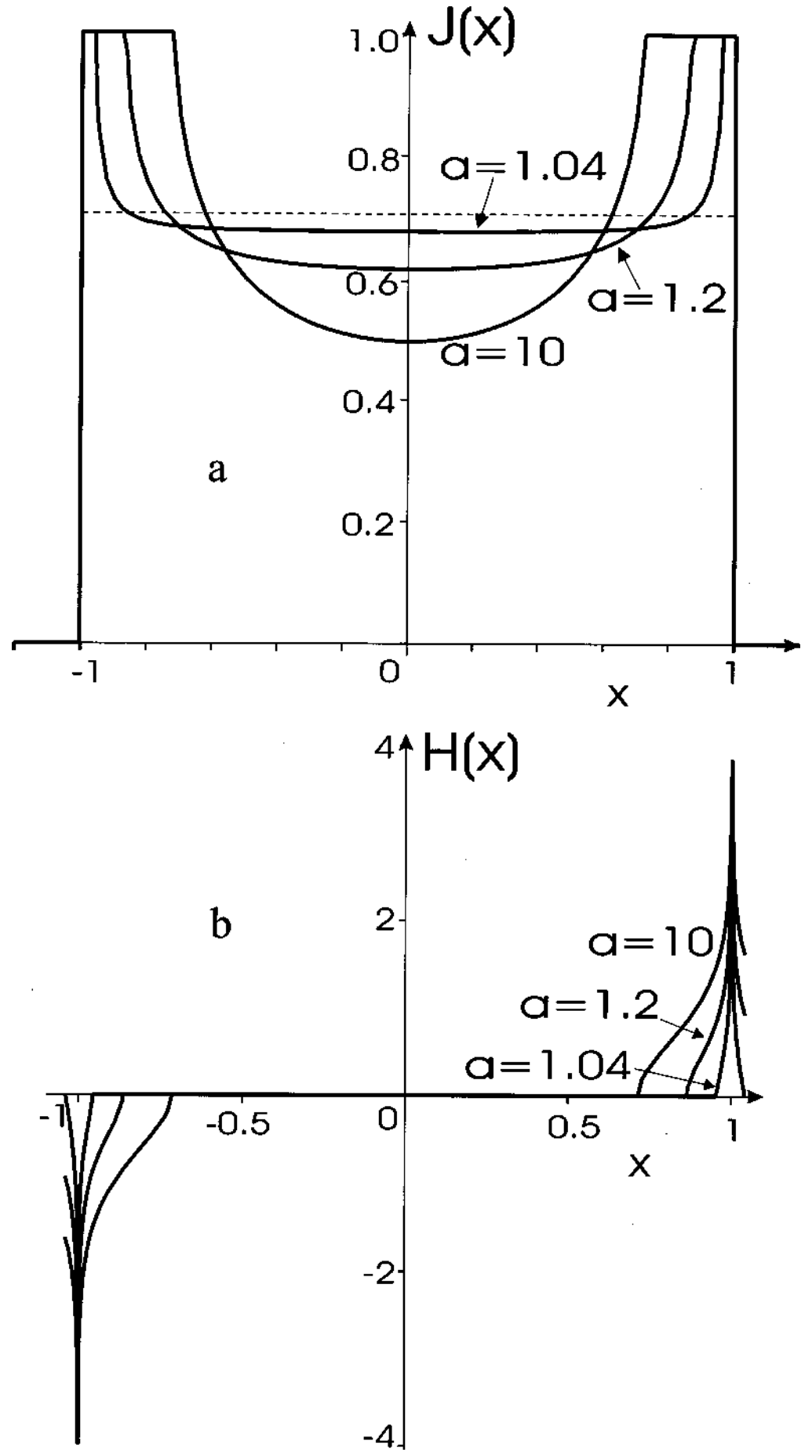

FIG. 16. Transport current (a) and perpendicular to the strip magnetic-field (b) distributions over the sheet carrying the total current $I=0.7\cdot I_c$ are shown for various distances $a$ from the sheet center to the magnets in perpendicular configuration. In the limit case $a=1$ the entire sheet is flux-free and sheet current $J=0.7\cdot J_c-\text{const}$ (dashed line).

$$J(x)=\begin{cases}\dfrac{2J_c}{\pi}\arctan\sqrt{\dfrac{l^2-p^2}{p^2-\left(\tan\dfrac{\pi}{2a}x\right)^2}}, & |x|<b,\\ J_c, & b<|x|<W/2.\end{cases} \tag{65}$$

A corresponding field distribution follows from Eq. (50) and reads as

$$H(x)=\begin{cases}0, & |x|<b,\\ \dfrac{H_c x}{|x|}\operatorname{arctanh}\sqrt{\dfrac{\left(\tan\dfrac{\pi}{2a}x\right)^2-p^2}{l^2-p^2}}, & b<|x|<\dfrac{W}{2},\\ \dfrac{H_c x}{|x|}\operatorname{arctanh}\sqrt{\dfrac{l^2-p^2}{\left(\tan\dfrac{\pi}{2a}x\right)^2-p^2}}, & |x|>\dfrac{W}{2}.\end{cases} \tag{66}$$

The width of the flux-free region $2b$ is defined by the value of the total current. This may be found by integration of the current distribution (65) which gives

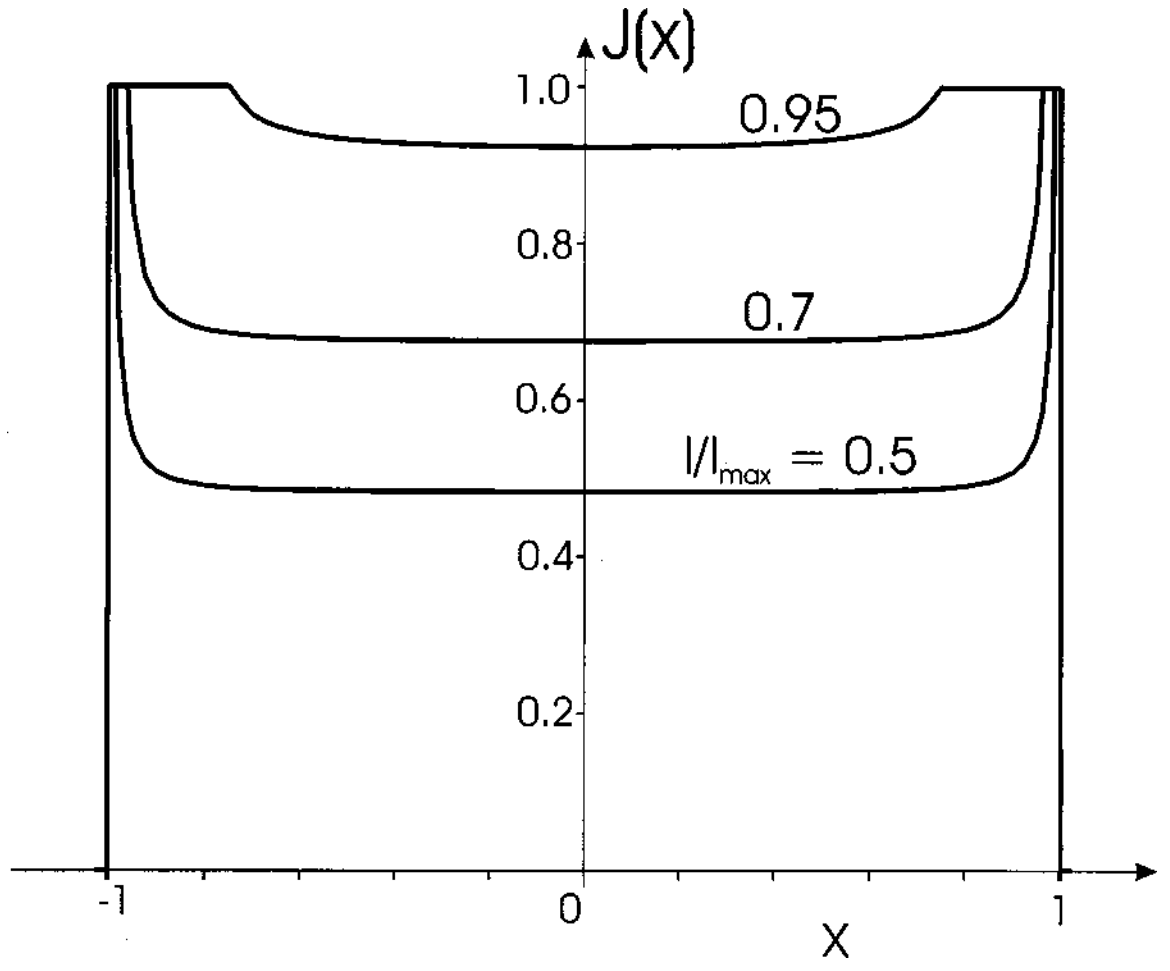

FIG. 17. Current distributions in the partly flux-filled state of a superconducting strip with total currents $I/I_c=0.5$, 0.7, 0.95, ($I_c=J_cW$) placed between magnetic shields perpendicular to the strip plane at a distance $a-1=0.04$ (in units of $W/2$).

$$p=\sqrt{\frac{\cos\left(\frac{\pi W}{2a}\frac{I}{I_c}\right)-\cos\frac{\pi W}{2a}}{1+\cos\frac{\pi W}{2a}}},$$

$$b=\frac{2a}{\pi}\arctan p. \qquad (67)$$

Sheet current distributions over the strip and field patterns of a fixed total current at various distances to the magnets $a$ are shown in Fig. 16. An approaching of the magnet walls to the strip causes a flattening of the current distribution in a flux-free region and an extension of the latter. The reasons for this flattening are physically transparent if one takes into account that, at $a\rightarrow W/2$, the strip with its images, as well as films array in the work of Mawatari,[32] form a continuous infinite superconductor strip lying on the $x$ axis. The Meissner state of the latter is provided by a constant sheet current.

Figure 17 presents an evolution of the current distribution for different total currents at the small fixed magnet-superconductor distance $2a/W=1.04$. The picture exhibits a close analogy to the results for the Meissner state of a film with a large edge barrier.[2] In fact, pinning in Fig. 18 acts like an extended edge barrier providing a wide flux-free zone inside a strip up to the largest magnitudes of a total current $I\simeq I_c$.

This analogy allows one to suppose that a special design of the magnetic surrounding may permit a redistribution of the current to the center of the strip similar to that obtained above in Sec. III for the edge barrier mechanism. We study this question in the next section.

## V. OVERCRITICAL STATES OF A HARD SUPERCONDUCTOR STRIP IN AN OPEN CURVED MAGNETIC CAVITY

As was established in Ref. 2 and above in Sec. III B the maximum nondissipative current flowing in a flux-free superconductor strip may be strongly enhanced in open curved magnetic cavities. It was shown also in the previous section that, in a partly flux-filled state, the central flux-free region is very sensitive to the magnetic surrounding too. That is why we study in this section how a current distribution behaves in a strip partly filled with a pinned magnetic flux and placed in a convex magnetic cavity.

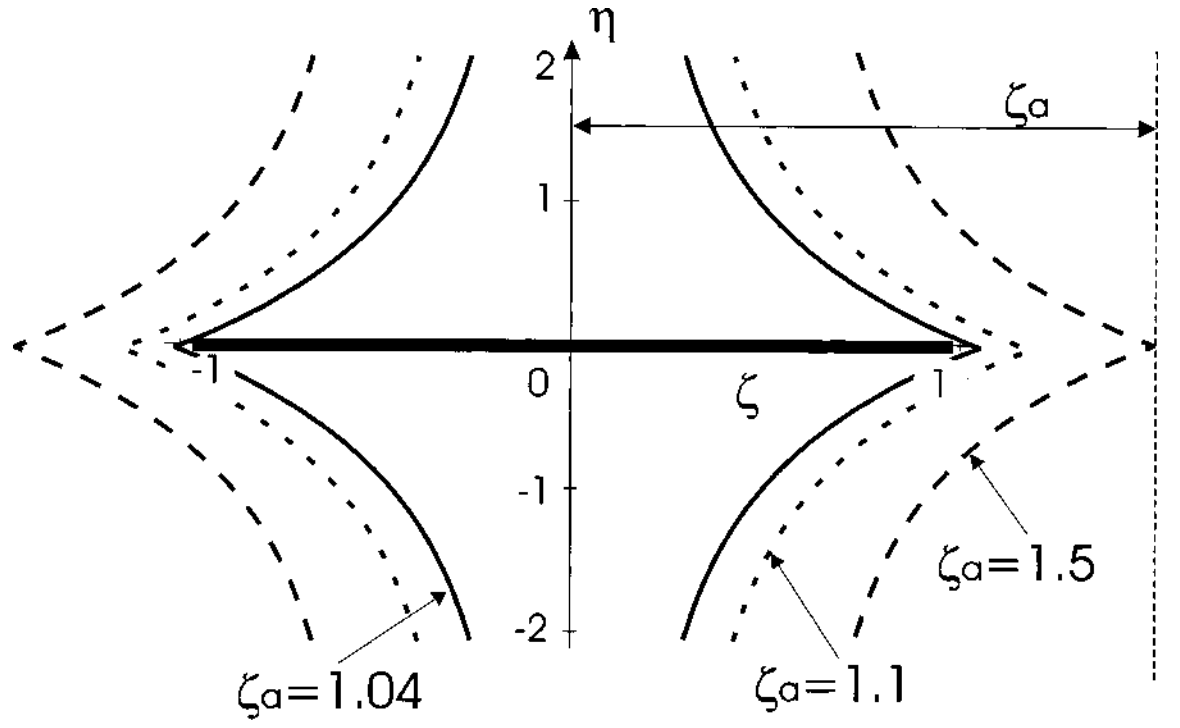

FIG. 18. Current-carrying strip inside an open magnetic cavity obtained by conformal mapping from configuration of Fig. 15 to a new frame $(\zeta,\eta)$ for various distances $\zeta_a$ from the strip center to the cavity vertex. The strip width in new coordinates is $2\zeta_W=2$.

For the study of the pure Meissner state of a strip surrounded by a magnet with $\mu\gg1$ the mighty tool of a conformal mapping was used in Sec. III B. In that case, magnetic-field lines are parallel to the strip everywhere at the strip surface and almost perpendicular to the surface of the magnet shields in analogy to an electrostatic field near the surface of a conductor. Conformal mapping keeps these properties of field lines during the deformation of the geometry and thus allows one to construct the Meissner state in various geometries if it is known in some configuration.

In the case of the partly flux-filled state, conformal mapping cannot be applied directly. Field lines in this case are still parallel to the strip in the flux-free region [see Eqs. (43)], which holds during the conformal mapping, but the second property, $J=J_c$ in the flux-filled region, does not hold generally. Thus the state resulting from the mapping of some partly flux-filled state does not have the form (43). Nevertheless, conformal mapping may be used for the constructing of partly flux-filled states in complicated geometries following the scheme advanced by Norris and summarized above in Sec. IV A.

In our problem, we seek first for the Meissner response of the flux-free region $|x|\leq b$ to the unit test currents located at $x=\pm x_0$, $x_0>b$ with account of the flat magnetic walls located at $x=\pm a$, $a>x_0$ (see perpendicular configuration in Sec. IV C). The obtained solution is transformed then by conformal mapping to the current distribution over some flux-free strip in the curved magnetic surrounding arising in response to external test currents. Finally, the partly flux-filled state in a curved surrounding is constructed by integration over the test current position in the constant-$J_c$ region. The integration of the resulting sheet current over the strip width gives the total current flowing in the strip. This way the relation between the main parameter of the problem, the flux-free zone width $2b$, and the total current is established.

A magnetic field induced by a current strip between magnetic walls perpendicular to it is given by Eqs. (63) accounting for multiple images of the strip. The equation for the Meissner response of the flux-free region to two test currents of unit strength flowing along the lines $x=\pm x_0$ is found by substitution of the form (44) into Eq. (64) which gives

$$\int_{-b}^{b} du\, J(u)\cot\left(\frac{\pi}{2a}(x-u)\right)=\cot\left(\frac{\pi}{2a}(x_0-x)\right)-\cot\left(\frac{\pi}{2a}(x_0+x)\right). \tag{68}$$

With the help of substitutions used in Sec. IV C and $t_0=\tan(\pi x_0/2a)$ one can reduce the above equation to the

$$\frac{1}{\pi}\int_{-p}^{p}\frac{d\tau\phi(\tau)}{\tau-t}=\frac{1+t_0^2}{2a}\left(\frac{1}{t+t_0}+\frac{1}{t-t_0}\right), \quad |t|\le p. \tag{69}$$

This equation reproduces, in principle, Eq. (45) and its solution follows with an adjusting factor from the solution (47) and reads

$$\phi_0(t)=-\frac{t_0(1+t_0^2)}{a(t_0^2-t^2)}\sqrt{\frac{t_0^2-p^2}{p^2-t^2}}. \tag{70}$$

One can check that the integration of $\phi_0(t)$ over the test current position $x_0$ in the interval $b<x_0<W/2$ gives the partly flux-filled state (65).

We are now in a position to perform a conformal mapping of a strip carrying a current

$$J_0(x)=\phi_0[t(x)]+\delta(x-x_0)+\delta(x+x_0), \tag{71}$$

where $t(x)=\tan \pi x/2a$, and located between magnets with $\mu\to\infty$ occupying the space at $|x|\ge a$ to another geometry.

Let us carry out the mapping of the walls $x=\pm a$ to the convex cavity similar to the transformation (36). This mapping of the initial complex plane $s=x+iy$ onto a new one $w=\zeta+i\eta$ is performed by an analytical function

$$w=f(s)=c[\sqrt{1+s/a}-\sqrt{1-s/a}], \tag{72}$$

where a factor $c=b[\sqrt{1+b/a}-\sqrt{1-b/a}]^{-1}$ is chosen to keep the position of the flux-free region: $f(\pm b)=\pm b$. Let us note that the strip half-width $W/2$ and distance to the magnets $a$ are changed after the mapping to the $\zeta_W=f(W/2)$ and $\zeta_a=f(a)$, respectively.

The form of magnetic walls resulting from the conformal mapping of flat walls (Fig. 15) is shown in Fig. 18 for various distances between the strip and magnets. Note that, to get different solutions for a given flux-free zone width and strip width, initial strips with different widths should be chosen for mapping.

The sheet current distribution over the strip after the mapping is given by a jump of a parallel field component as is presented by Eq. (31). Then the test current $\Delta J=\delta(x-x_0)$ transforms to

$$\Delta\tilde{J}(\zeta)=\left.\frac{\partial y}{\partial\eta}\right|_{\eta=0}\delta[x(\zeta,\eta=0)-x_0]=\left.\frac{\partial y}{\partial\eta}\frac{\delta(\zeta-\zeta_0)}{|\partial x/\partial\zeta|}\right|_{\eta=0}, \tag{73}$$

where $\zeta_0=f(x_0)$. Making use of Eqs. (37) and (72), one finds

$$\left.\frac{\partial y}{\partial\eta}\right|_{\eta=0}=\left.\frac{\partial x}{\partial\zeta}\right|_{\eta=0}=\frac{a}{c}\frac{1-\zeta^2/2c^2}{\sqrt{1-\zeta^2/4c^2}}. \tag{74}$$

Thus, in new coordinates, the test currents remain currents of unit strength.

A Meissner response current acquires then a form

$$\tilde{J}_M(\zeta)=\frac{a}{c}\frac{1-\zeta^2/2c^2}{\sqrt{1-\zeta^2/4c^2}}\phi_0\{t[x(\zeta,\eta=0)]\}, \tag{75}$$

where $x(\zeta,\eta=0)=(a\zeta/c)\sqrt{1-\zeta^2/4c^2}$.

This solution presents a current flowing in the region $|\zeta|\le b$ and shielding this region from a field induced by the two test currents of unit strength with account of contribution from surrounding curved magnets. In the new coordinates, the test currents are flowing along the lines $\zeta=\pm\zeta_0$ and magnet surfaces assume positions determined parametrically as

$$w=c[\sqrt{1\pm 1+iy/a}-\sqrt{1\mp 1-iy/a}] \tag{76}$$

with $-\infty<y<\infty$ as is shown in Fig. 18.

A current distribution for the partly flux-filled strip in the curved magnetic surrounding may be obtained by integration of the solution (75) over $\zeta_0$ in the constant-$J_c$ region $(b,\zeta_W)$. As well as after integration of Eq. (47), a contribution arises which diverges at the edges of flux-free region $\zeta=\pm b$. Since the solution to the inhomogeneous Eq. (69) includes a solution to a corresponding homogeneous equation with an arbitrary factor [see also Eq. (46)], this unphysical divergence may be canceled by adding of a conformally mapped solution of the homogeneous Eq. (69) with a proper factor. The latter Meissner solution which presents a flux-free state of a strip $|x|\le b$ placed in the curved magnetic cavity defined by the mapping (72) is given by Eq. (38) where one should substitute $c$ for $c_a$ and $b$ for $W/2$.

After this regularization a continuous current distribution looks finally as

$$\tilde{J}(\zeta)=\frac{J_c}{\sqrt{2}\pi}\frac{1-\zeta^2/2c^2}{\sqrt{1-\zeta^2/4c^2}}\int_0^{\varphi_0}d\varphi\left\{\left[\frac{\pi}{2}+\arctan(\sqrt{p^2+(p^2-t^2)(\tan\varphi)^2})\right]^{-1/2}+\left[\frac{\pi}{2}-\arctan(\sqrt{p^2+(p^2-t^2)(\tan\varphi)^2})\right]^{-1/2}\right\} \tag{77}$$

when $|\zeta|<b$ and $\tilde{J}(x)=J_c$ when $b<|\zeta|<\zeta_W$. In the above formula,

$$t=\tan\left(\frac{\pi}{2c}\zeta\sqrt{1-\zeta^2/4c^2}\right) \quad \text{and} \quad \varphi_0=\arctan\sqrt{\frac{l^2-p^2}{p^2-t^2}}.$$

Current distributions computed using the formula (77) are presented in Fig. 19. The initial strip width $W$ was chosen for every initial distance $a$ to flat magnets so that it is mapped onto the strip of a given width $2\zeta_W = 2$ while the flux-free region occupies the position $|\zeta| \leqslant b = 0.8$.

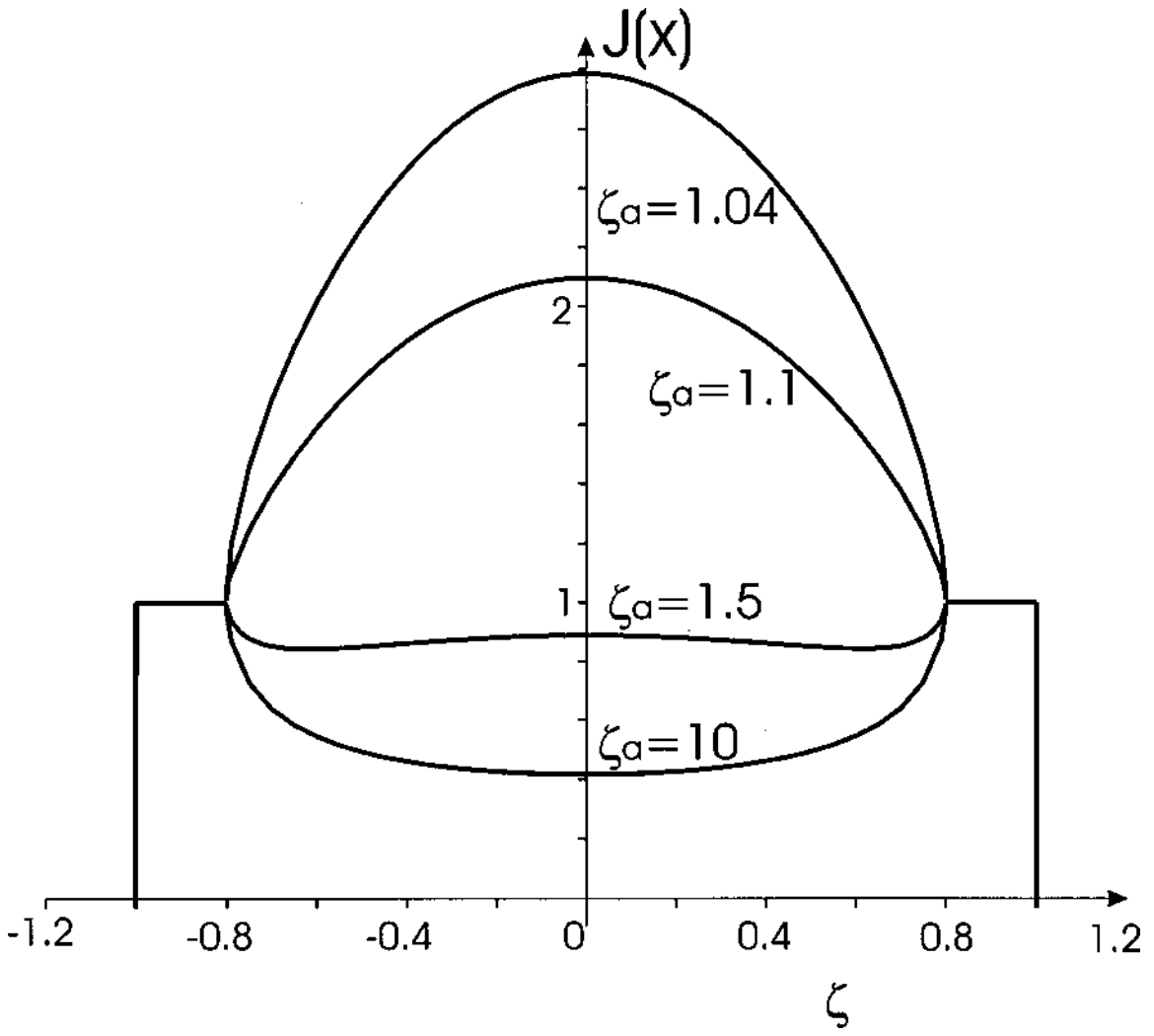

FIG. 19. Magnetically screened current-carrying state of a superconducting sheet of width $2\zeta_W = 2$ inside a cavity shown in Fig. 18. Meissner flux-free state occupies the region $|\zeta| < b = 0.8$. Different current profiles correspond to different distances $\zeta_a$ from the cavity vertex to the sheet center.

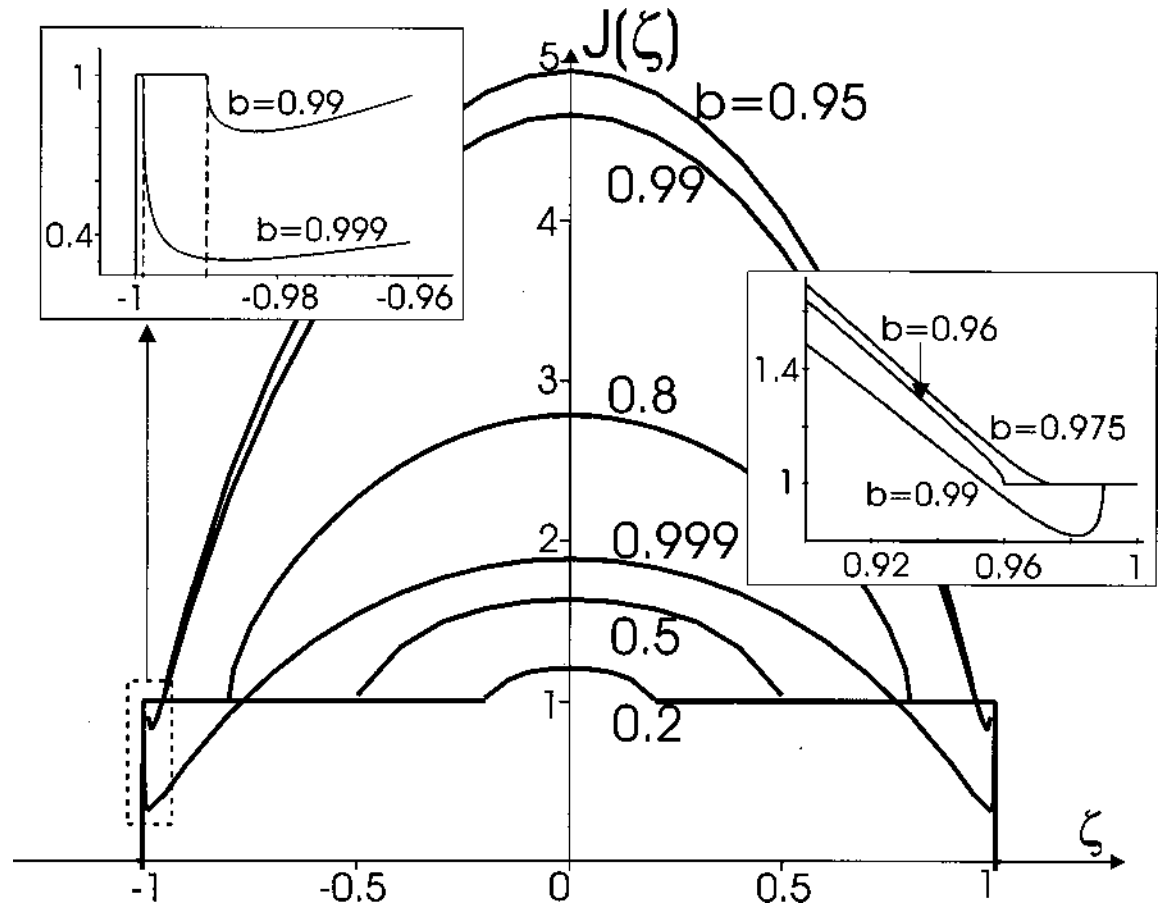

FIG. 20. Evolution of the sheet current distribution in a strip placed in a cavity shown in Fig. 18 with the change of the total current. The distance from the sheet center to the cavity vertex $\zeta_a = 1.04$ is fixed. Different curves correspond to different magnitudes of the total current which determine the different values of the width $2b$ of the flux-free zone. In the left inset, marginal parts of current distributions with narrow flux-filled regions are zoomed in. Right inset shows the transformation of the stable current pattern in unstable when parameter $b$ falls below the critical value of $b_0 = 0.975$.

One can observe that at distances $\zeta_a \gg \zeta_W (a \gg W)$ the current distribution reproduces the pattern characteristic of the isolated strip [Eq. (48) and Fig. 16]. When the magnets approach the strip to a distance $\zeta_a \simeq \zeta_W (a \simeq W/2)$ the current starts to redistribute to the center of the strip. At distances to the strip edge $\zeta_a - \zeta_W \ll \zeta_W$ the current distribution in the central flux-free region is inflated high above the critical current level thus providing a total transport current larger than that of the isolated strip in the critical state. However, the overcritical current distributions for $\zeta_a = 1.04$ and 1.1 turn out to be unstable with respect to transition to the common critical state with $\tilde{J} = J_c -$const all over the strip. Indeed, the vortices at the boundaries of the flux-filled region at $\zeta = \pm b$ would be dragged inwards by current $\tilde{J} > J_c$ until they meet in the center of the strip.

It is typical of hard superconductors that the current and field distributions are history dependent and therefore a state with a given transport current may be realized in many ways.[12,13] In the case of a magnetically screened strip the picture becomes even more complicated since, contrary to the isolated strip case,[10,12,13] the total current turns out to be a nonmonotonous function of the width of the flux-free region. To study the possibility of stable overcritical states we consider now the evolution of the current distribution in the strip located at a small fixed distance to magnets $\zeta_a = 1.04$ while gradual increase of the total transport current from zero.

At small total currents, the constant-$J_c$ domain $b < |x| < \zeta_W$ grows but remains first very small so that the whole strip but narrow margins is in the flux-free state. By further increase of the current it concentrates mostly in the strip center where it may exceed $J_c$. At the total current $I = 2.74 I_c$, where $I_c = 2 J_c \zeta_W$ is the total critical current of the isolated strip, only $10^{-3}$ of the strip is penetrated by flux ($b = 0.999$) as is seen in the left inset in Fig. 20. Note the vertical slope of current distributions at boundaries of the flux-free region which is typical of two-dimensional flux and current distributions.[10,12,13]

For $b = 0.99$ and 0.999 the region with the overcritical current $J > J_c$ is separated from the flux-filled margins by undercritical regions with $J < J_c$ which prevents the flux-front propagation inwards. Thus the corresponding current distributions are stable. On the contrary, the current patterns with $b = 0.95$ and smaller in Fig. 20 exhibit a direct contact of the flux-filled regions with the overcritical region which makes these distributions unstable.

Computation results presented in the right inset in Fig. 20 reveal the critical value of the flux-free zone width $b = b_0 = 0.975$ when the undercritical region first vanishes. The corresponding current distribution lies above the other distributions and thus delivers the largest total current of $I_{max} = 7.3 I_c$. This is the maximum stable current which can be carried by the strip in the magnetic cavity of Fig. 18 with $\mu \gg 1$ and distance $\zeta_a - \zeta_W = 0.04 \cdot \zeta_W$ between magnets and superconductor. The distributions with $b < b_0$ carry smaller currents and are unstable with respect to transition in usual critical state with $\tilde{J} = J_c -$const.

Note that current distribution derivatives at $\zeta = \pm b$ are infinite but of different signs for $b > b_0$ and $< b_0$. This allows one to find the critical value $b_0$ from the condition $J'(\zeta \to \pm b) = 0$. We consider for simplicity the case of the small distance magnet superconductor $\zeta_a / \zeta_W = 1 + \epsilon$, $\epsilon \ll 1$ which gives $W/2a \simeq 1 - 2\epsilon^2$ and $l = \tan(\pi W/4a) \simeq 1/\pi\epsilon^2 \gg 1$. It is also assumed that the flux-filled region of interest is narrow so that $W/2 - b \ll W/2$ which gives $p = \tan(\pi b/2a) \simeq 2a/\pi(a-b) \gg 1$. In this limit, the current distribution (77) may be simplified to

$$\tilde{J}(\zeta) = \frac{J_c}{\sqrt{2\pi}} \frac{1 - \zeta^2/2c^2}{\sqrt{1 - \zeta^2/4c^2}} \int_0^{\varphi_0} d\varphi [p^2 + (p^2 - t^2)(\tan\varphi)^2]^{1/4}. \quad (78)$$

The main contribution in the derivative when $\zeta \to b-0$ (and hence $t \to p-0$) reads

$$\frac{\partial \tilde{J}}{\partial \zeta} \sim \frac{J_c}{\sqrt{p^2-t^2}} \left( 1+2\sqrt{\frac{l}{p}} - \frac{l^2}{p^2} \right) \tag{79}$$

which exhibits the singularity at $t \to p$ in qualitative agreement with Fig. 20. A polynomial with respect to $\sqrt{l/p}$ in brackets has the only positive root $h=1.395$ which provides the criterion of a stable current distribution:

$$\tan\left( \frac{\pi b}{2a} \right) \geqslant h^{-2} \tan\left( \frac{\pi W}{4a} \right). \tag{80}$$

For the strip of unit half-width $\zeta_W = 1$ and $\epsilon \ll 1$ this results approximately in

$$b_0 = 1 - \epsilon (h-1). \tag{81}$$

This simplified expression gives $b_0 = 0.984$ for $\epsilon = 0.04$ used above in Fig. 20 which differs by 10% from the value $b_0 = 0.975$ found by computation of exact Eq. (77).

Taking into account that the integral in Eq. (78) depends weakly on $\zeta$ one can estimate the maximum total current as

$$I_{max} = I_c \sqrt{\frac{1}{4\pi} \tan\left( \frac{\pi W}{4a} \right)} \simeq \frac{I_c}{2\pi\epsilon}, \quad a - W/2 \ll W/2. \tag{82}$$

This means formally that the total current grows infinitely as $a \to W/2$ ($\zeta_a \to \zeta_W$). In fact, the usage of the formulas (80) and (82) is restricted by the smallest reasonable superconductor-magnet distance $\zeta_a - \zeta_W \simeq \delta$ (which is followed by $a - W/2 \simeq 4\delta^2/W$) since at smaller distances our basic Eqs. (3), (5), and (8) are no longer valid. Practical reasons impose much stronger restrictions on this distance which seems to be at best of the order of 0.1 mm. This gives, for the film of width 5 mm, an enhanced total current of $I_{max} \simeq 4 \cdot I_c$ which is less than enhancement 7.3 found above numerically. Thus approximate expressions (78)–(82) give underestimated values of the total current. Nevertheless, using formula (82) one finds, for the least physically reasonable distance of $\delta/W \simeq 10^{-3}$, the huge enhancement of the total current $I_{max} \simeq 160 \cdot I_c$.

## VI. DISCUSSION AND CONCLUSIONS

In this paper we studied theoretically the completely flux-free Meissner state and partly flux-filled critical state of current-carrying superconductor strips placed in a magnetic surrounding. No external field or remanent magnetization was implied. Only the self-field of a transport current was considered with account of surrounding magnetic medium which is supposed to be homogeneous, linear, and reversible. Analytical results were obtained for field and current distributions over the strip in the limit of the infinite magnetic permeability of magnets $\mu \to \infty$ for various shapes of magnet boundaries.

The analytical solutions of singular integral equations describing a current distribution were found for the flux-free sheet located near a flat bulk magnet surface and inside open wedge-shaped cavities for the case of an infinite magnetic permeability $\mu \to \infty$. A strong suppression of current peaks at the sheet edges and the redistribution of the current to the inner part of the sheet were shown to strongly enhance a current-carrying capability of the sheet in the Meissner state.

Numerical studies for finite $\mu$ allow one to conclude that the current peak height at the sheet edges depends practically linearly on the parameter $q = (\mu - 1)/(\mu + 1)$ and 90% of the suppression effect is reached already for $\mu \simeq 30$. The current redistribution thanks to the magnetic screening by $\mu > 100$ is represented to an accuracy of 1% by the analytical solutions obtained for $\mu \to \infty$.

Physically important for all above solutions is that the sheet remains in the flux-free Meissner state. Then the redistribution of the current is not accompanied by the motion of magnetic vortices and may not be prevented by pinning. The Meissner state of the sheet is supposed to be protected by the large edge barrier against the magnetic flux entry and may be distinctly observed only on samples whose critical current is essentially controlled by the barrier effect. Thanks to the edge barrier and the current redistribution over the sheet induced by a magnetic surroundings it turns out to be possible to carry a high total transport current.

The most crucial parameter for the substantial current redistribution is the distance between the bulk magnets and the superconducting sheet. The direct contact ($a \to 1$) and a consequently complete suppression of the current peaks at the edges is hardly achievable. Practically plausible gaps between the magnet surfaces and the sheet are about 0.1 mm which gives, for the sheet width $W = 5$ mm, the least possible value of the distance to the sheet center $a = 1.04 \cdot W/2$. For the most robust geometrical barrier mechanism, this gives an average current in the Meissner state of about $10^6$ A/cm$^2$ in a flat magnetic surroundings (Fig. 15) and a current about $10^7$ A/cm$^2$ in a convex magnetic cavity for a YBCO film of thickness $d = 50$ nm at 77 K (Figs. 11 and 12).

Hard superconductors possessing strong pinning and negligible edge barrier exhibit similar behavior in the partly flux-filled state. We found that the current distribution in the central flux-free part of the strip is very sensitive to the shape of the magnetic shields and their distance to the strip. In a simple case of flat magnet surfaces the current behaves exactly as in the stack and linear array of superconductor films[32] for strip positions parallel and perpendicular to the magnet walls, respectively.

In the parallel configuration, the current is expelled from the flux-free central region reminding, at close contact of the magnets to the strip, the well-known Bean's field and current profiles for a bulk slab.[1,26] In the perpendicular configuration, on the contrary, the current distribution in the flux-free zone is flattened. At close contact of the strip to the magnets the major part of the strip remains flux-free even at high total currents $I \simeq I_c$, the critical current of the isolated strip, thus reminding the current distribution in a magnetically screened superconducting film with a high edge barrier against flux entry.

The above results apply to the virgin state when a transport current grows from zero. An arbitrary succession of current switching (on and off) produces more complicated history-dependent current and field patterns arising because of the pinning-induced irreversibility. These patterns in the isolated strip studied in details in Refs. 12 and 13 may be simply reproduced for the strip in a flat magnetic surrounding by proper substitutions of variables[32] which make the problems identical.

The most unusual behavior of the current distribution was found for the strip placed in an open convex magnetic cavity. Contrary to the current concentration near slab surfaces and strip edges typical of flux-free superconductors, the current is redistributed in this case to the center of the flux-free zone where it can exceed the pinning-determined sheet critical current $J_c$.

The filling of the strip with a transport current starts, at small total currents, with a very nonmonotonous distribution which has some extrema in a flux-free region. As in usual critical state,[1,26] the sheet current is constant and equal $J_c$ in the flux-filled margins along the strip edges. These margins are as narrow as the distance magnet-superconductor and become infinitesimal simultaneously with the latter. By further increase of the total current the sheet current in the flux-free zone grows and becomes finally larger than $J_c$. The constant-$J_c$ regions remain very small thus reminding the current behavior of strips with a high edge barrier.[2] The difference from the latter is that, even at finite distance to magnets, there are no edge peaks in the current distribution.

The overcritical flux-free region where the sheet current $J>J_c$ is first separated from the flux-filled margins with $J=J_c$ by undercritical flux-free regions where $J<J_c$ which secure the overcritical state. At some value of the total current $I_{max}$ the undercritical regions vanish which let the current $J>J_c$ directly contact the vortices at the boundary of the flux-filled margins that makes this configuration unstable. Then the vortices dragged by current $J>J_c$ move inwards until they meet in the center of the strip thus restoring the usual critical state with $J=J_c$ all over the sample.

The maximum total nondissipative current $I_{max}$ which may be carried by the strip grows, in the considered configuration, approximately proportional to the inverse magnet-superconductor distance. This distance is restricted from below by some characteristic length equal to or exceeding the film thickness, which leads to $I_{max}$ values two orders of magnitude larger than $I_c$. Practically, the magnet-superconductor distance can hardly be made smaller than 0.1 mm which results in an $I_{max} \simeq 7 \cdot I_c$ for a film of width 5 mm.

For both flux-free strips secured by large edge barriers and hard superconductor strips dominated by pinning, other mechanisms restricting the total current may become valid. At high enough $J_c$ or $J_b$, the enhanced current in the strip center may reach the magnitude of the Ginzburg-Landau depairing current[25] and switch on a direct depairing process. In wide films, where current densities $j_c \simeq 10^6$ A/cm$^2$ where achieved at 77 K,[20] this may happen by total currents enhanced by one order over $I_c$.

Another more practical restriction is connected with the violation of the used one-dimensional scheme when magnetic vortices enter from the top and bottom surfaces of the strip where magnetic field exceeds the lower critical field parallel to strip $H_{c1}^{\parallel}$. To avoid this process, thin films with a thickness $d \ll \lambda$ may be used[2] where the parallel critical field is enhanced up to $(\lambda/d)^2 H_{c1}^{\parallel}$.[37] This should not present a problem for YBCO films possessing a $\lambda=0.5$ $\mu$m at 77 K.[19]

Other experimental difficulties may be connected with the magnetic shielding material which should possess, at low temperatures, a high enough $\mu$ and a high saturation field (to be linear at $H \sim J_c$), and which should behave reversibly, i.e., exhibit no remanent magnetization. The size of the bulk magnets may turn out to be crucial too. Since the magnetic field of the strip decreases very slowly with the distance inside the magnets and all calculations in the theory were performed for infinite medium, the size of the magnets should be much larger than the film width. Thus numerical calculations for finite-size and finite-$\mu$ magnets are necessary to establish exactly a possible current enhancement effect under real conditions.